# A Cloud-Based Energy Management Strategy for Hybrid Electric City Bus Considering Real-Time Passenger Load Prediction

Junzhe Shi[1], Bin Xu[2]*, Xingyu Zhou[3], Jun Hou[4]*

1: University of California, Berkeley, Department of Civil and Environmental Engineering, 760 Davis Hall, Berkeley, CA 94720, USA.

2: Clemson University, Department of Automotive Engineering, 4 Research Dr., Greenville, SC, 29607, USA

3: University of Texas, Austin, Department of Mechanical Engineering, 204 E. Dean Keeton Street, Stop C2200 ETC II 5.160 Austin, TX, 78712, USA.

4: University of Michigan, Department of Electrical Engineering and Computer Science, 1301 Beal Avenue, Ann

Arbor, MI, 48109, USA.

* Corresponding author: Bin Xu, Clemson University, Department of Automotive Engineering, 4 Research Dr., Greenville, SC, 29607, USA. Phone: 864-626-8335, Email: xbin@clemson.edu. Jun Hou, University of Michigan, Department of Electrical Engineering and Computer Science, 221 Santa Clara Way, San Mateo, CA 94403, USA. Phone: 734-272-5516, Email: junhou@umich.edu.

**Abstract - Electric city bus gains popularity in recent years for its low greenhouse gas emission, low noise level, etc. Different from a passenger car, the weight of a city bus varies significantly with different amounts of onboard passengers. After analyzing the importance of battery aging and passenger load effects on an optimal energy management strategy, this study introduces the passenger load prediction into the hybrid-electric city buses energy management problem, which is not well studied in the existing literature. The average model, Decision Tree, Gradient Boost Decision Tree, and Neural Networks models are compared in the passenger load prediction. The Gradient Boost Decision Tree model is selected due to its best accuracy and high stability. Given the predicted passenger load, a dynamic programming algorithm determines the optimal power demand for supercapacitor and battery by optimizing the battery aging and energy usage leveraging cloud techniques. Then, rule extraction is conducted on dynamic programming results, and the rule is real-time loaded to the vehicle onboard controller to handle prediction errors and uncertainties. The proposed cloud-based Dynamic Programming and rule extraction framework with the passenger load prediction show 4% and 11% lower bus operating costs in off-peak and peak hours, respectively. The operating cost by the proposed framework is less than 1% of the dynamic programming with the true passenger load information.**

## 1. Introduction

### 1.1 Background

Greenhouse gas emissions by human activities have been dominating global warming since the mid-20th century [1]. Reducing carbon dioxide emissions by minimizing conventional fossil fuel applications is the key to sustainable development for the whole world. In particular, from the transportation perspective, the electrified propulsion systems [2] has become a promising alternative to the conventional internal combustion powertrains across the globe.

The hybrid energy storage system (HESS), which combines batteries and supercapacitors, has high potentials in vehicular applications because it entails the advantages of both supercapacitors and batteries, including high power density and high energy density [3]. Among different types of vehicles, city buses are ideal for applying the HESS to reduce carbon dioxide emissions due to the high passenger loading capacities [4]. The performance of a complex HESS highly depends on its energy management strategy (EMS) [5]. The manufacturing cost of a hybrid electric city bus is mostly on the batteries, which have a much shorter cycle life than supercapacitors [6]. Thus, the EMS of hybrid electric city buses, aiming to minimize both electric consumption cost and battery degradation cost during operation, is worth being investigated.

1.2 Literature Review

In the literature, various EMSs have been designed and discussed for hybrid electric buses. The most popular methods include the rule-based method, Dynamic Programming (DP), and Model Predictive Control (MPC).

The rule-based method is commonly exploited to devise the EMS in the existing automotive industry for real-time applications, and it builds rules based on expert knowledge or heuristic principles. A rule-based EMS, rooted in DP optimization, is proposed for an electric bus with battery aging considered [7]. In the simulation, operation cost improvements are obtained, and such a rule-based strategy achieves a battery life extension. A fuzzy-logic rule-based EMS is proposed to determine series and parallel operation mode for a hybrid electric bus [8]. The results show the proposed rule-based method effectively controls the energy consumption of the engine and maintains battery charge levels. Simulation results indicate less energy consumption than that of conventional engine-only drivetrain. However, these rules are fixed once they are developed and do not adapt to real-time varying conditions.

Different from rule-based methods, DP and MPC are optimization-based methods. DP is a global optimization method that can address system nonlinearities and constraints. [9] utilizes a DP to deal with the integrated optimization for deriving energy splitting strategies of a hybrid electric city bus. A preset cost function is employed to evaluate the HESS life cycle cost, and derive the best configuration and energy split strategies of a hybrid electric city bus system, including a battery and a supercapacitor. MPC is an optimal control strategy that is used to solve optimization problems with a designed predication horizon while satisfying a set of constraints. DP can also be used as an optimization solver for MPC, but its high computational cost limits the its real-time applications for MPC. In [10], an MPC with a Pareto-front analysis of the objective functions is used to reduce fuel consumption and battery degradation. However, MPC only considers short prediction horizons and its solution is only local optimal. As discussed in the abovementioned papers, optimization-based methods provide more desired cost-saving results than rule-based methods do. However, these optimization-based methods require accurate predictions of future operating conditions; once driving conditions are different from the predicted one, the performances of the above methods cannot be guaranteed. Besides, the high computational burden and real-time feasibility of DP and MPC are always concerns.

Nowadays, the rapid development of intelligent transportation systems and the widespread use of cloud computing offer excellent opportunities to improve the overall performance of energy management strategies [11]. However, to the best of authors' knowledge, only limited literature discuss the methods of using cloud techniques to improve the EMS in hybrid electric vehicles. [12] applies both MPC and DP in a plug-in hybrid electric bus, and provides a feasible solution to address the high computational cost of DP and collect future power demand data by utilizing vehicle-to-cloud (V2C) connectivity. A two-layer data-driven supervisory EMS which leverages V2C connectivity is proposed to increase energy efficiency [13]. In the EMS, a DP is run in the cloud layer to minimize energy consumption, while an MPC scheme is operated in onboard layers to solve the real-time powertrain control problem. However, this EMS does not consider battery degradation in the optimization. Besides, the above two EMSs require a high computational MPC running at the vehicle level, and do not guarantee the real-time feasibility of MPC.

Unlike a normal passenger car, the weight change of a city bus due to the passenger load effect can be up to 40% of its total weight [14]. By taking the parameters of the newest Mercedes-Benz eCitaro into count, the passenger load effect can even lead to 43% of the total weight change of the bus [15]. Power demand varies significantly along with different passenger load when the buses are travelling at relatively high speed and accelerations [16]. For example, a higher passenger load leads to a higher power demand peak during acceleration and requires more control efforts to avoid charging or discharging batteries with high C-rates to minimize the battery capacity loss. An EMS of hybrid electric city buses, which does not consider the passenger load effect, will lead to extra operation cost and thus should not be considered as an optimal control strategy. However, there is a paucity of literature that discusses the passenger load effect on an optimal energy management strategy generation of hybrid city buses.

In summary, the following research gaps are identified based on the literature review:

1) Rule-based methods lack optimization. MPC only considers short prediction horizon and its solution is only local optimal. Even though DP generates global optimal solutions, it is a computationally heavy algorithm and can only be executed offline. It is challenging to execute DP in real-time energy management.

2) Only a limited amount of literature exploits cloud techniques and proposes a practical method to improve the EMS of hybrid electric vehicles.

3) Existing bus energy management literature does not systematically consider passenger load variation, which leads to a huge gap between algorithmic optimized result and actual optimal solution.

### 1.3 Motivation and Contribution

Vehicle-to-Infrastructure (V2I) and Vehicle-to-Vehicle (V2V) hold a significant potential to improve the information acquisition for the prediction of future traffic information [17]. Besides, Vehicle-to-Cloud (V2C) provides a chance to store and process the massive amount of data, update the structures and parameters of current algorithms, and apply the accurate prediction algorithms and optimal control algorithms that are both computationally expensive. With the powerful cloud cluster, even the DP algorithm will have the potential to execute in real-time. One of the main purposes of this study is to propose a comprehensive framework of hybrid electric city buses EMS, which fully explores and utilizes the benefits of cloud techniques.

This paper extends the existing research by discussing the effect of the passenger load and battery aging on the optimal EMS generation. The simulation results show that the passenger loading effect plays a more dominant role than the battery aging effect does on the optimal control strategy generation. Due to several stochastic variables (e.g., weather conditions), a precise passenger load prediction is challenging [18]. There are several works focused on passenger demand prediction on bus networks. [19] propose an interactive multiple model-based pattern hybrid approach to predict short-term passenger demand using historical data and real-time observations with 24.3% absolute percentage error. Besides, three city bus passenger demand prediction models, including time-varying Poisson model, autoregressive integrated moving average model, and sliding window ensemble framework model, are presented in [20] and achieve 78% prediction accuracy. To further improve the passenger load prediction accuracy of existing literature, this paper investigates the method of passenger load prediction using different statistics and machine learning models. Because the inputs (i.e., weather condition, time-of-day, day-of-week, etc.) are not highly corelated, the supervised models shine in these complex modeling problems. This paper also proposes a novel hierarchical EMS of hybrid electric city buses to explore the benefits of V2C and V2I connectivity and cloud computing.

The schematic diagram of using cloud data collection and computation for hybrid-electric buses is presented in Fig. 1. At the cloud level, the prediction of the reference power demand is achieved by collecting the traffic information and predicting passenger load. Then, the planning of the reference power splitting for hybrid electric city buses is generated with the concept of "block MPC" on the cloud by DP method. Running the DP on the cloud solves a highly nonlinear costs function, provides an optimal solution, and saves the computational cost at the vehicle level. On the vehicle side, a rule-based controller with rule update ability is used to handle the error between the predicted and actual power demands with the limited computational cost. In summary, the contributions of this study are as follows:

1) Battery aging effect and passenger load effect on an optimal energy management strategy generation of hybrid city buses are analyzed by using a DP approach with supercapacitor and battery electric models, a Li-ion cycle-life battery degradation model, and an electric bus dynamics model.

2) Passenger load prediction is introduced into the hybrid-electric city buses energy management problem. Four models are investigated for the passenger load prediction, including an average model and three supervised learning models. The contributions of input features to the passenger load prediction are analyzed based on feature importance ranking. As the on-board passenger number varies significantly during rush hours and off-peak hours, the passenger load prediction substantially increases the accuracy of power demand profile prediction for EV buses.

3) A cloud-based EMS of hybrid electric city buses considering real-time passenger load prediction is proposed to take advantage of the V2I and V2C connectivity. The performance of minimizing electric consumption cost, battery degradation cost, and the low computational cost on the bus side is guaranteed.

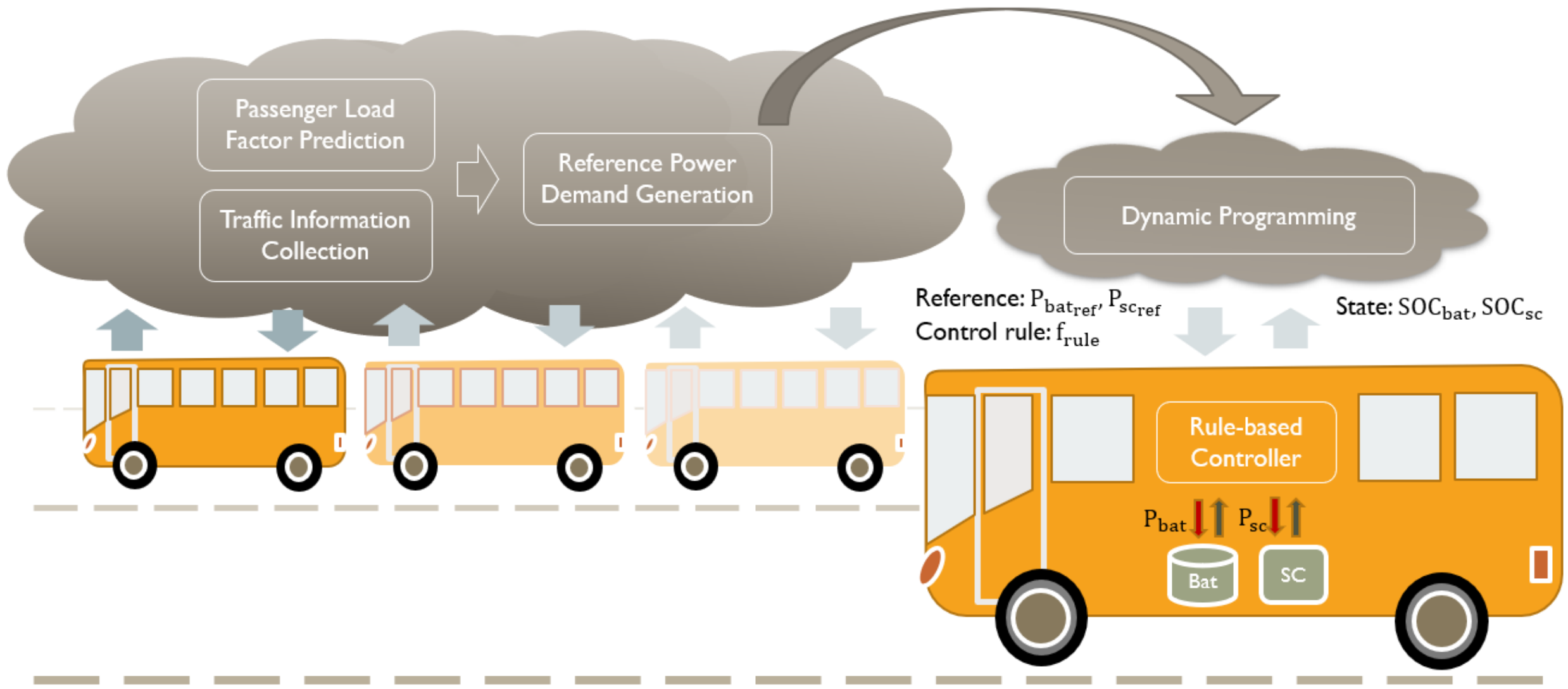

Fig. 1. The schematic diagram of cloud data collection and computation for hybrid-electric buses.

This paper is organized as follows. Section 2 introduces the system modeling and configuration. Section 3 presents methodologies about the passenger load prediction models and EMS. Section 4 first analyzes the aging effect and passenger loading effect, then presents the results of different passenger loading prediction models, compares the three different EMS, and discusses the adapting of the proposed method into wider applications in the end. Section 5 highlights the key conclusions of this paper.

## 2. Vehicle Configuration and Modeling

In this section, the vehicle dynamics and system configurations of a hybrid electric city bus are presented. Besides, the models of battery and supercapacitor used in this study are discussed.

### 2.1 Vehicle Model and System Configurations

According to the specification of Mercedes-Benz eCitaro [15], the maximum energy capacity, maximum load capacity, and empty vehicle mass of the vehicle model are determined. The model aims to convert speed profiles to power demand profiles for the hybrid energy storage system based on different driving conditions. The vehicle dynamics can be expressed as,

$$m_{whole} g f v \cos(\alpha) + 0.5 C_D A_f \rho v^3 + \frac{m_{whole} v dv}{dt} + m_{whole} g v \sin(\alpha) = \begin{cases} P_{demand}\eta_T\eta_{md}, & P_{demand} > 0 \\ \dfrac{P_{demand}}{\eta_r}, & P_{demand} \le 0 \end{cases} \quad (1)$$

$$m_{whole} = m_{vehicle} + m_{person} C_{max} f_{load} \quad (2)$$

where v is the vehicle speed, α is the road slop, $f_{load}$ is the passenger load factor, and the rest of the parameters are summarized in Table 1.

Table 1

Parameters and Description of the Electric City Bus [15].

| Parameter | Description | Value |
|---|---|---|
| $m_{vehicle}$ | Empty vehicle mass | 13500 kg |
| $m_{person}$ | Average mass of a person | 70 kg |
| $C_{max}$ | Maximum passenger load capacity | 145 |
| $E_{max}$ | maximum energy capacity | 293 kWh |
| g | Gravity acceleration | $9.8\ m/s^2$ |
| $A_f$ | Front area | $7.5\ m^2$ |
| f | Rolling resistance coefficient | 0.018 |
| $C_D$ | Air drag coefficient | 0.7 |
| ρ | Air density | $1.29\ kg/m^3$ |
| $\eta_T$ | Transmission efficiency | 90% |
| $\eta_{md}$ | Electrical machine efficiency | 85% |
| $\eta_r$ | Regenerative braking efficiency | 65% |

According to the required energy capacity of Mercedes-Benz eCitaro, the battery configuration is set up as 217 in series and 7 in parallel in this study. To supply at least 15 seconds of peak power consumption, the supercapacitor pack is set up as 20 in series and 6 in parallel. The topology of the semi-active hybrid energy storage system is presented in Fig. 2. The electric motor drives the rear wheels. A DC/AC converter locates between the electric motor and the two energy storage systems. A DC/DC converter is used to mitigate the voltage among battery, supercapacitor, and AC/DC converter.

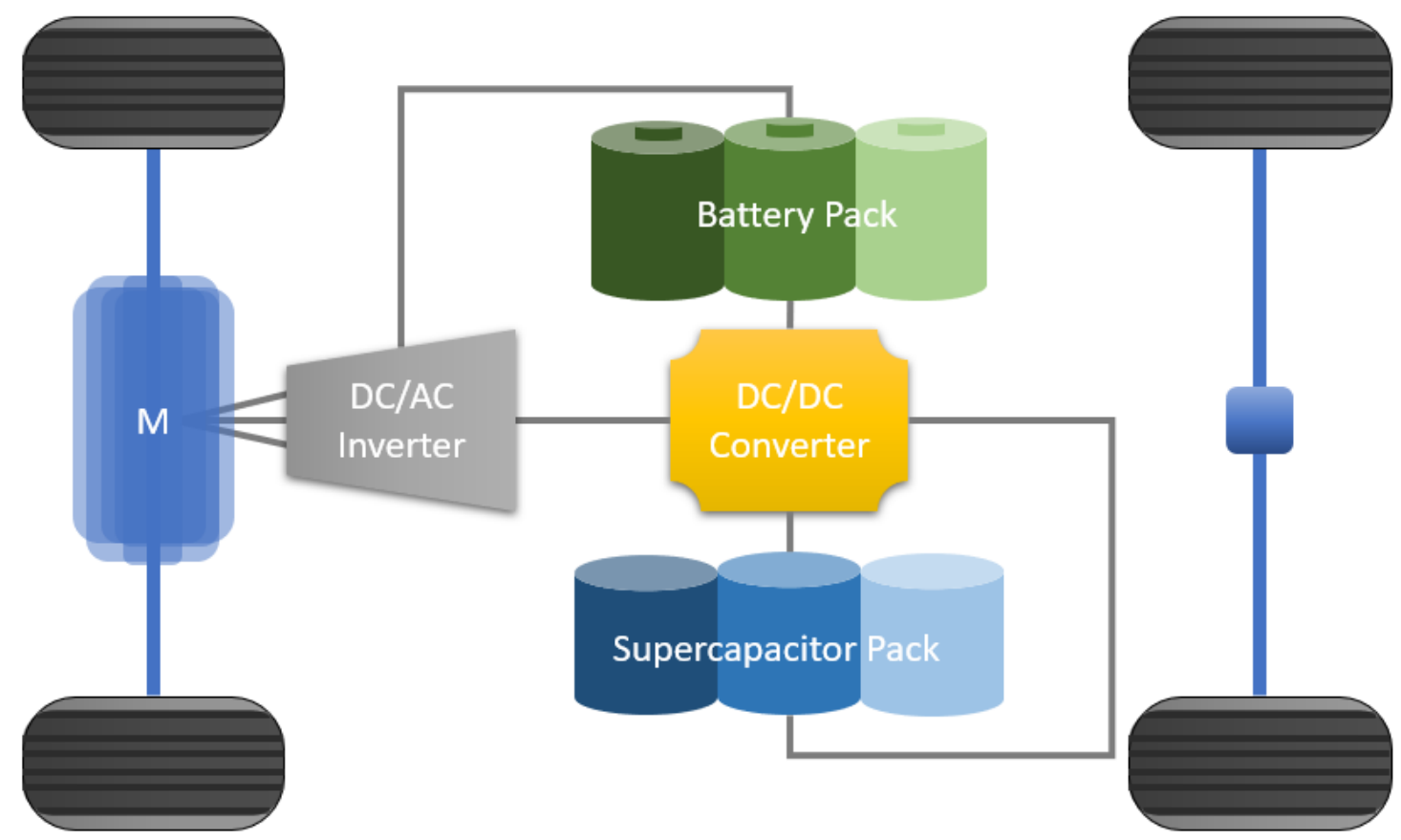

Fig. 2. Topology schematic of a semi-active hybrid energy storage system.

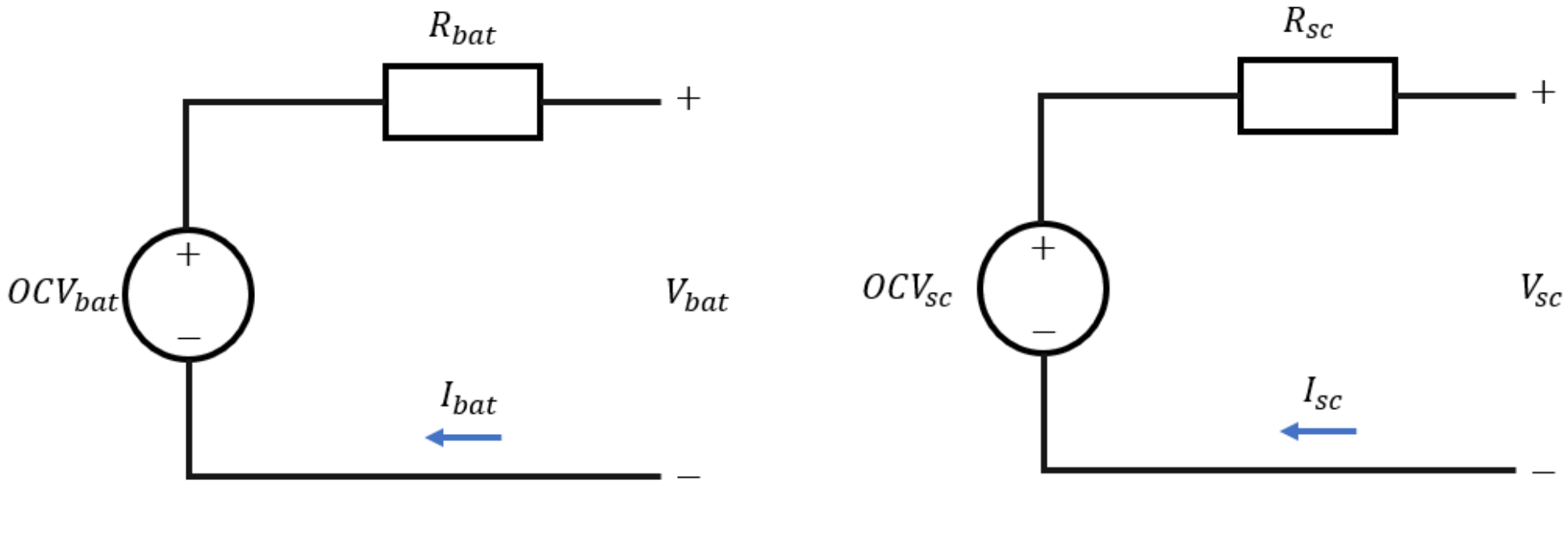

(a) Rint model of a battery

(b) RC model of a supercapacitor

Fig. 3. Equivalent circle models of a battery and supercapacitor.

2.2 Battery Model

In this study, the state-of-charge (SOC) of a battery is defined as a state variable, while the current of the battery is defined as the control input. The dynamic of a battery is,

$$\dot{SOC}_{bat} = -\frac{I_{bat}}{3600Q_{bat}} \tag{3}$$

where the $Q_{bat}$ is the nominal capacity of a battery.

As shown in Fig. 3 (a), the equivalent circle model of a battery comprises an open-circuit voltage (OCV) and an ohmic resistor. The output and electric loss powers of a battery can be calculated by using,

$$P_{bat_{loss}} = I_{bat}^2 * R_{bat} \quad (4)$$

$$P_{bat_{output}} = OCV_{bat} * I_{bat} - I_{bat}^2 * R_{bat} \quad (5)$$

where $OCV_{bat}$ and $R_{bat}$ are the OCV and the ohmic resistance of a battery. Their values depend on the $SOC_{bat}$, and the other parameters of the battery used in this study are listed in Table 2.

Table 2

Battery Parameters [21].

| Parameter | Value |
|---|---|
| Capacity (Ah) | 60 |
| Stored Energy (kWh) | 0.192 |
| SOC Usage Window (%) | 10 - 90 |

The capacity of a battery limits the maximum energy that a battery can supply. However, the capacity of a battery will fade as the battery ages. Different aging models can be found in literature [22]. Through cycling tests, an empiric aging model is obtained and verified in [21]. The capacity fade depends strongly on battery usage, C-rate, operating temperature, and depth-of-discharge. The battery depths-of-discharge (operating range) of a hybrid electric city bus is assumed to range from 10% to 90% SOC. The semi-empirical life model adopts the following equation to describe the correlation between the capacity loss, Ah throughput, temperature, and C-rate,

$$Q_{loss} = 0.0032 * \exp\left(-\frac{15162 - 1516 * C_{rate}}{RT_{bat}}\right) * Ah^{0.824} \quad (6)$$

where the $Q_{loss}$ represents the normalized capacity degradation, R is the gas constant equal to 8.3145 $\frac{J}{molK}$, $T_{bat}$ is the absolute temperature of the battery, and Ah is the Ah throughput. After discretizing the formula of the battery degradation equation Eq. (6), a dynamic degradation model is obtained,

$$\Delta Q_{loss} = \left(0.0032 * \exp\left(-\frac{15162 - 1516 * C_{rate}}{RT_{bat}}\right)\right)^{\frac{1}{0.824}} * Q_{loss}^{\frac{0.824-1}{0.824}} * 0.824 * \frac{\Delta t * I_{bat}}{3600} \quad (7)$$

where $\Delta t$ is the sampling time. While a battery is aging, the value of its resistance and capacity will change. In this study, the resistance and capacity of the batteries are assumed to be obtained and updated by online estimation.

2.3 Supercapacitor Model

As the same as a battery, the SOC of a supercapacitor is also defined as a state variable, and the current of a supercapacitor is defined as the control input,

$$\dot{SOC}_{sc} = -\frac{I_{sc}}{C_{sc} V_{max\,sc}} \quad (8)$$

where the $C_{sc}$ and $V_{max\,sc}$ are the capacitance and maximum voltage of a supercapacitor.

As shown in Fig. 3 (b), the Rint-Capacity model of a supercapacitor is comprised by an OCV and an ohmic resistor. The OCV of a supercapacitor is determined by,

$$OCV_{sc} = SOC_{sc} * V_{max\,sc} \tag{9}$$

The output and electric loss powers of a supercapacitor can be calculated by using,

$$P_{sc_{loss}} = I_{sc}^2 * R_{sc} \tag{10}$$

$$P_{sc_{output}} = OCV_{sc} * I_{sc} - I_{sc}^2 * R_{sc} \tag{11}$$

where $R_{sc}$ is the resistance depending on the C rate, and the other parameters of the supercapacitor used in this study are listed in Table 3.

Table 3

Supercapacitor Parameters [21].

| Parameter | Value |
|---|---|
| Max Voltage (V) | 27 |
| Capacity (F) | 140 |
| Stored Energy (kWh) | 0.0142 |
| SOC Usage Window (%) | 50 -100 |

Unlike a battery, a supercapacitor has a much longer cycle life, and its degradation is neglectable [23]. Therefore, the supercapacitor cycle life mode is not considered in this study.

## 3. Methodology

### 3.1 Passenger Load Factor Prediction

A Passenger Load Factor (PLF) represents the capacity utilization of public transport services. 0% PLF means the bus is empty, while a 100% PLF means the bus is filled. By applying Eq. (2), the whole mass of the city bus can be updated basing a predicted PLF.

This section focuses on discussing different models of the PLF predictions, including an average model and three regression models - Neural Network (NN), Regression Tree (RT), and Gradient Boosting Decision Tree (GBDT).

3.1.1 Passenger Load Data Analysis

The actual passenger load data was collected from line 11 city buses in Guangdong, China [24]. The raw data recodes the total number of people who took the line 11 city buses each hour from 08/01/2014 to 12/24/2014, as shown in the blue line of Fig. 4. There are four spike regions each month, which represents the weekdays busy hours of the four weeks. Each spike region has five lines, representing the five weekdays.

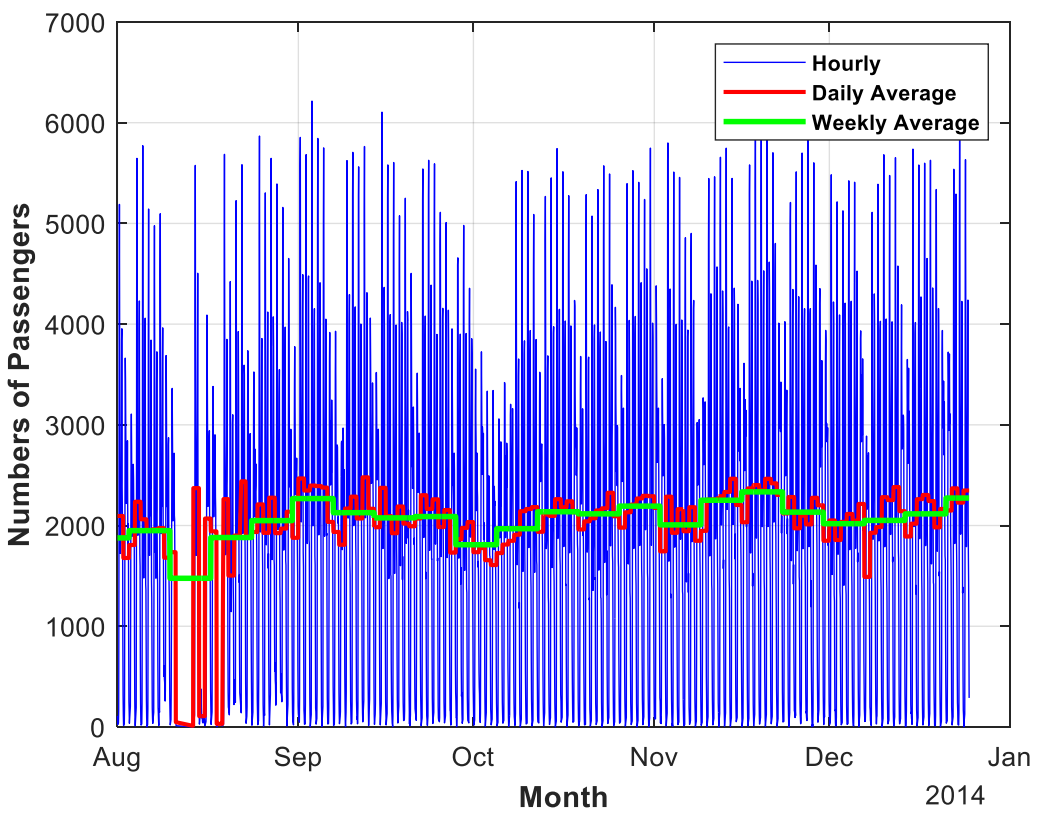

Fig. 4. The number of passengers who took the line 11 city buses in Guangdong from 08/01/2014 to 12/24/2014.

3.1.2 Average Model

The hourly, daily averages, and weekly averages of the passenger load data are also shown in Fig. 4. The pattern of daily average data is more regular than the weekly average data. Thus, the passenger load data is normalized and classified according to the day of a week. As shown in Fig. 5, the blue lines represented the PLF of a particular day, and the red lines indicate hourly average PLF of all weeks. Besides, the day-of-week data patterns can also work as the average model to predict the PLF in the future. The average model well represents the trend of PLF on a particular day of a week. From Monday to Friday, the peaks of the passenger loading happen at 8:00 AM and 6:00 PM, which are exactly the commuting time of most people. The average number of people who take buses during the weekend is less than the numbers on the weekdays.

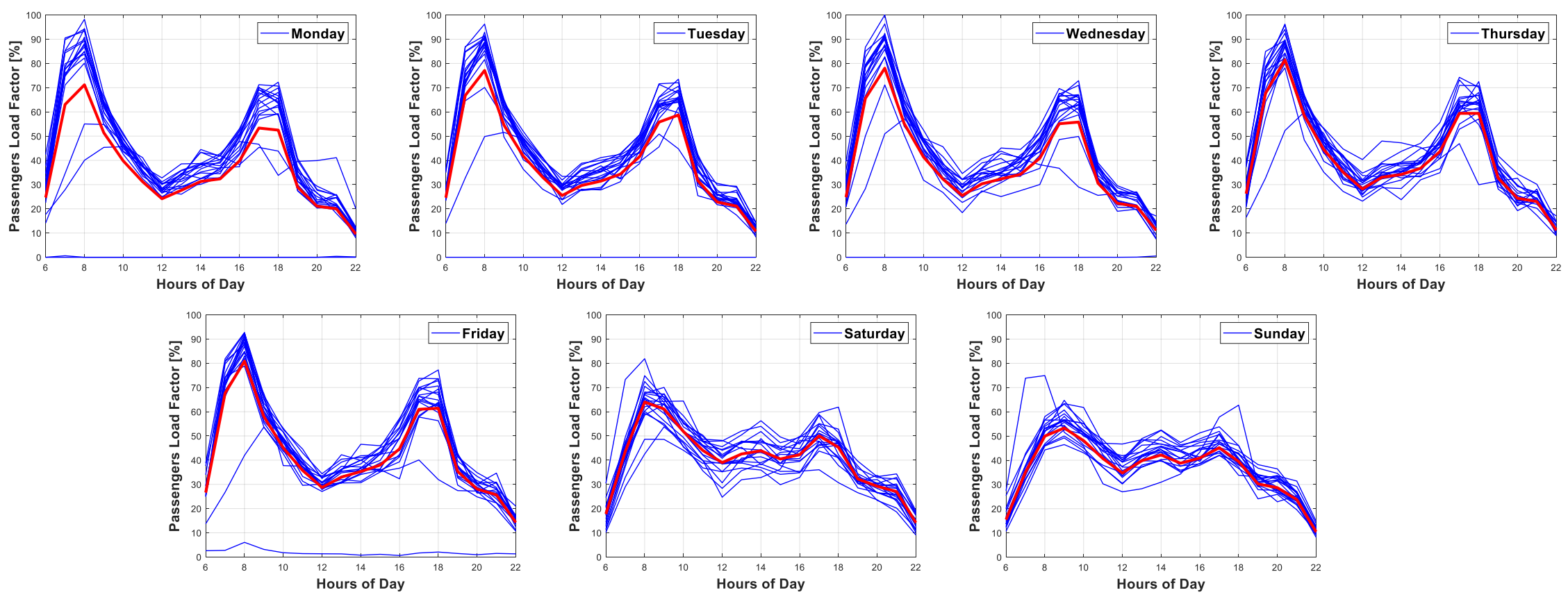

Fig. 5. Passengers load Factor from Monday to Sunday. Blue lines: passenger load of a particular day. The red line: hourly average passenger load of all weeks.

### 3.1.3 Regression Models

In this paper, the regression models, including Neural Network, Regression Tree, and Gradient Boosting Decision Tree, are considered and evaluated to increase the prediction accuracy of the average model. The inputs of these regression models are day-of-week, time-of-day, weathers, temperatures, wind levels, and holiday information. The data of temperatures is presented in Fig. 6. The structures of each regression method are described in the following subsections.

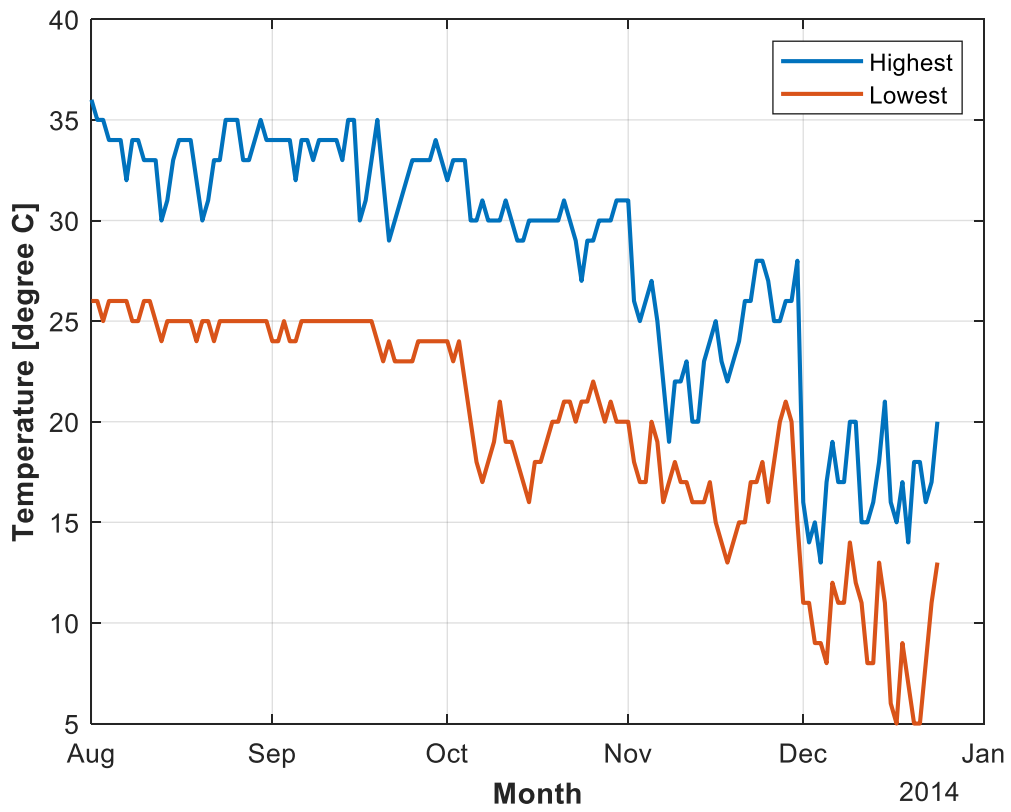

Fig. 6. The temperatures in Guangdong from 08/01/2014 to 12/24/2014. The blue line: the highest temperatures each day. The red line: the lowest temperatures each day.

#### 3.1.3.1 Neural Network

Neural Networks (NN) are a class of statistical learning algorithms. Fig. 7. illustrates the structure of the four layers perceptron network used in the study.

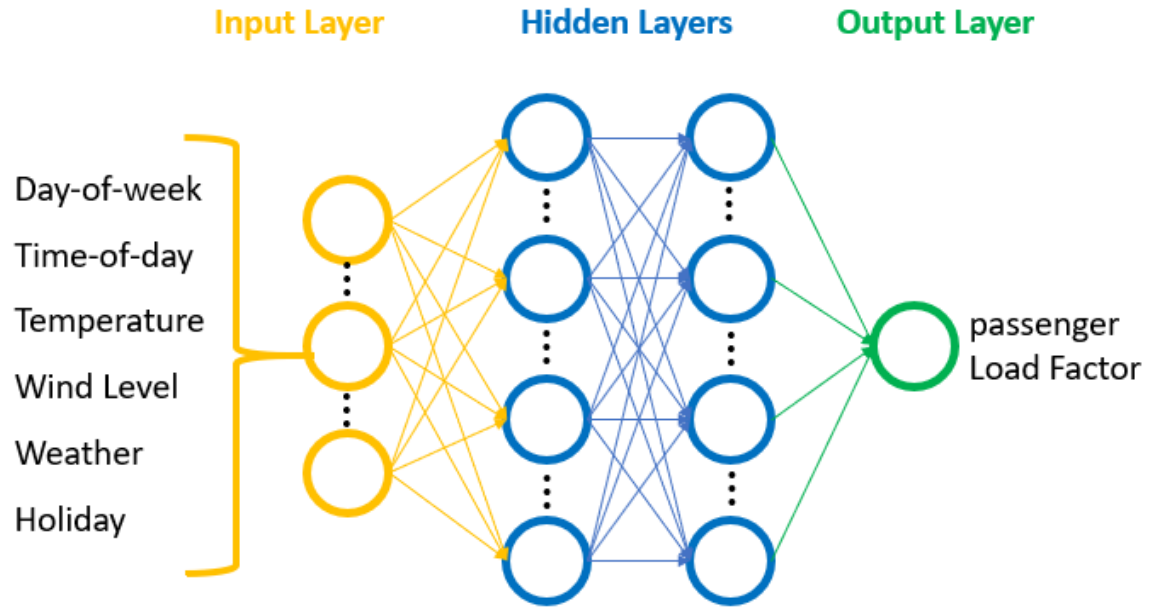

Fig. 7. Neural Network structure for the PLF prediction.

The hidden layer performs nonlinear transforms for the feature extractions. Because the ReLu function provides fast calculation and converging speeds without a gradient vanishing issue, in the hidden layers, the ReLu function is selected as the activation function,

$$\mathrm{ReLU} = \max(0, \mathrm{x}) \tag{12}$$

For the neuron in the output layer, a linear activation function is applied.

During the training, the backpropagation algorithm is used to adjusts the weights and thresholds of each neuron to minimize the error between the true PLF and predicted results of the NN with current parameters sets up. In this paper, the NN is trained using gradient descent backpropagation for 200 epochs. For a further discussion of neural networks, readers can find more details in [25].

3.1.3.2 Regression Tree

Regression Tree (RT) is one type of decision tree and a machine learning method for predictions. Fig. 8. illustrates the structure of the Regression Tree used in the study. As a tree structure, the feature space of the RT is recursively partitioned into small parts to create its node and leaves. Once the splitting is completed, each leaf is fitted by a simple regression model.

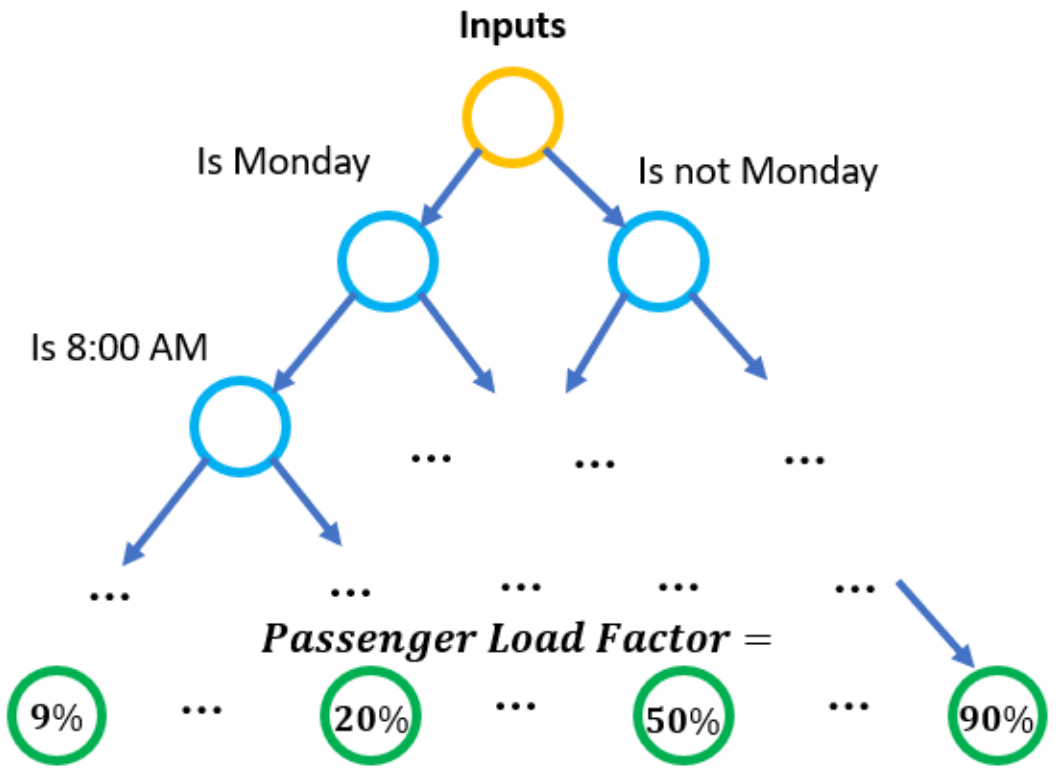

Fig. 8. Regression Tree structure for the passenger load factor prediction.

During the training process, the tree recursively splits a node (data set) into two child nodes (subsets). By exhaustively searching over data space, the division, composed of feature and threshold, is selected to minimizes the total impurity of two child nodes. In the paper, the impurity is defined as the sum of squared deviations between each PLF data and the mean of all the PLF variables in this branch, $\bar{y}$. For a branch with N variables, the impurity functions P is given by,

$$\bar{y} = \frac{1}{N}\sum_{i=1}^{N} y_i \tag{13}$$

$$P = \frac{1}{N}\sum_{i=1}^{N} (y_i - \bar{y})^2 \tag{14}$$

The training process stops after the relative decrease in the impurity is below a prespecified threshold, or the maximum tree depth is reached. For more discussions about a Classification and Regression Tree, more details can be found in [26].

3.1.3.3 Gradient Boosting Decision Tree

A Gradient Boosting Decision Tree (GBDT) uses an ensemble of decision tresses for predictions. During the training, a series of small decision trees are built. Each decision tree attempts to correct residuals from the previous one.

In this paper, the mean of the PLF of the whole data set is selected as the baseline of the predictions. A decision tree with a shallow max depth can be constructed to predict the residuals based on the previous prediction results. With the additional decision tree, the prediction with inputs is updated. After prediction values are updated, residuals are calculated again to generate a new decision tree, which further improves prediction accuracy. The training processes repeat until the maximum numbers of decision trees are created. Once trained, all of the decision trees in the ensemble are used to make a final prediction by,

$$\hat{y}(X_i) = b + \sum_{m=1}^{N} \alpha * DT_m(X_i) \tag{15}$$

where b is this baseline, α is the learning rate that controls how hard each new tree tries to correct remaining residuals from the previous round, DT(X) represents the predicted residual with given inputs, and N is total numbers of decision trees. Readers can find more details about the Gradient Boosting Decision in [27].

### 3.2. Energy Management Strategy

As shown in Fig. 9, in this section, an innovative energy management strategy, combining cloud technologies, dynamic programming, and rule-based controller with the rule update, is proposed for hybrid-electric buses. The frame and flowchart of the proposed method are demonstrated in Fig. 9. With the assistance of cloud computing, real-time traffic information can be collected, and the algorithm mentioned in the previous section about the prediction of PLF can be implemented. Taking the PLF and traffic information as inputs, the vehicle dynamics model described in Section 2.1 is used to generate a reference power demand profile. By taking advantage of the power cloud cluster, with the inputs as the reference power demand and states of batteries and supercapacitors, DP is executed to generate reference power profiles of batteries and supercapacitors. A control rule extraction is real-time conducted, and the rule is loaded to the onboard vehicle controller via the V2I connectivity. On the vehicle side, besides applying the reference power profiles of batteries and supercapacitors generated by the cloud DP, a rule-based controller with rule update ability is used to handle the error between the reference and actual power demands.

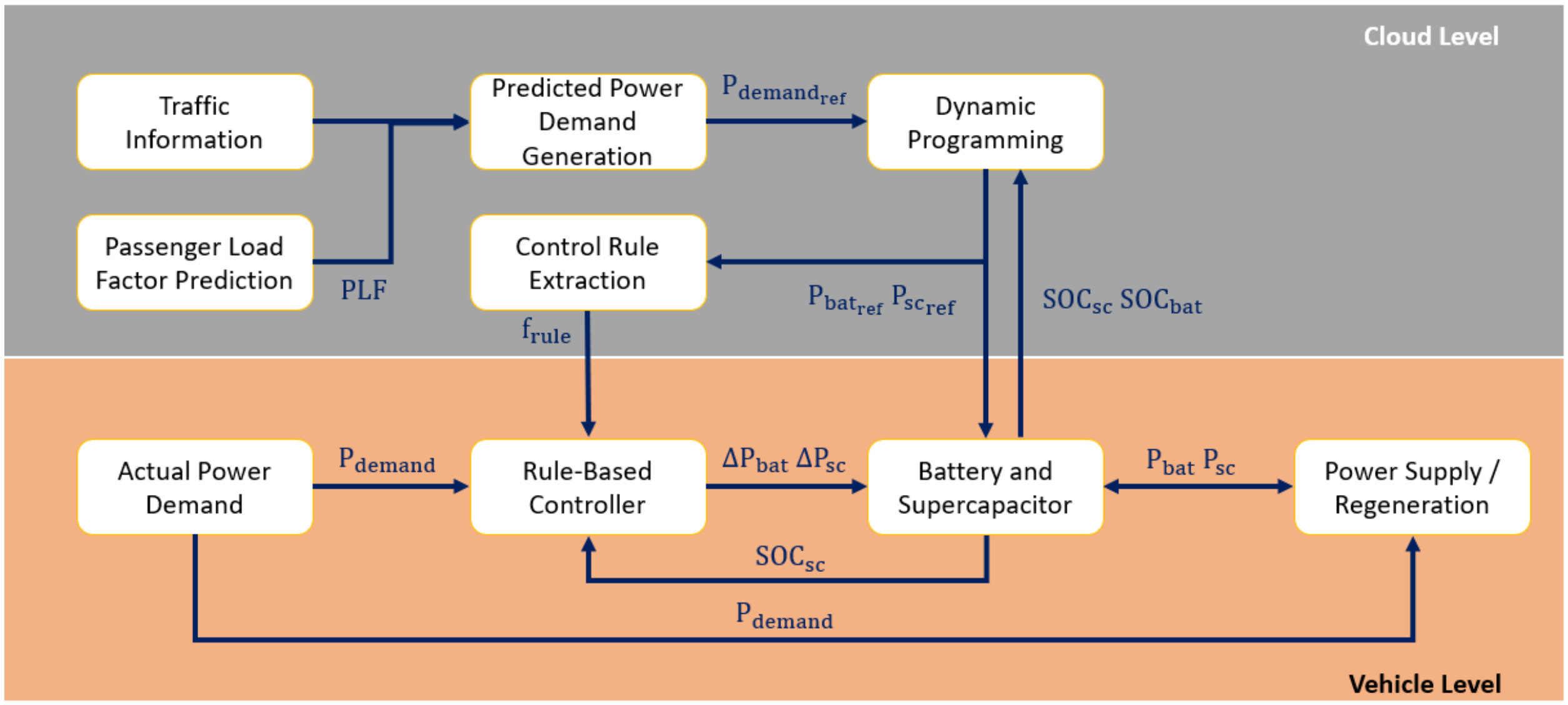

Fig. 9. Frame and flowchart of the proposed EMS.

#### 3.2.1 Dynamic Programming through Cloud Computing

First proposed by Richard Bellman in 1950s [28], the dynamic programming method breaks down a complex optimization problem into multiple subproblems, solves the subproblems, saves the solutions into a table, and then finds the solutions for the complex problem by looking up the table. The advantages of dynamic programming include: 1) providing a globally optimal solution, 2) capability for multi-stage decision processes, 3) applicability of nonlinear systems, and 4) handling of constraints.

In this study, the cost function of dynamic programming is formulated as follows:

$$J(k) = \sum_{t=k}^{N} \left( \Delta Q_{loss}(k) * C_{bat} + \Delta E_{loss}(k) * C_{ele} + W_{sc} * \sigma_{sc}(k) + W_{bat} * \sigma_{bat}(k) \right) \tag{16}$$

subject to the system dynamics:

$$SOC_{bat}(k+1) = SOC_{bat}(k) + \frac{-T_s I_{bat}(k)}{3600 Q_{bat}} \tag{17}$$

$$SOC_{sc}(k+1) = SOC_{sc}(k) + \frac{-T_s I_{sc}(k)}{V_{sc_{MAX}} C_{sc}} \tag{18}$$

input constraints:

$$P_{demand_{ref}}(k) = P_{bat_{ref}}(k) + P_{sc_{ref}}(k) \tag{19}$$

$$P_{bat_{ref}}(k) = OCV_{bat}(k) I_{bat}(k) - R_{bat} I_{bat}(k)^2 \tag{20}$$

$$P_{sc_{ref}}(k) = OCV_{sc}(k) I_{sc}(k) - R_{sc} I_{sc}(k)^2 \tag{21}$$

$$I_{sc_{MIN}} \le I_{sc}(k) \le I_{sc_{MAX}} \tag{22}$$

$$I_{bat_{MIN}} \le I_{bat}(k) \le I_{bat_{MAX}} \tag{23}$$

state constraints:

$$SOC_{sc_{min}} - \sigma_{sc_{low}}(k) \le SOC_{sc}(k) \le SOC_{sc_{max}} + \sigma_{sc_{high}}(k) \tag{24}$$

$$SOC_{bat_{min}} - \sigma_{bat_{low}}(k) \le SOC_{bat}(k) \le SOC_{bat_{max}} + \sigma_{bat_{high}}(k) \tag{25}$$

$$\sigma_{bat} \ge 0 \tag{26}$$

$$\sigma_{sc} \geq 0 \tag{27}$$

and cost constraints:

$$\Delta E_{loss}(k) = \left(I_{bat}(k)^2 * R_{bat_{ser}} + I_{sc}(k)^2 * R_{sc_{ser}}\right) * T_s \tag{28}$$

$$\Delta Q_{loss}(k) = \left(0.0032 * \exp\left(-\frac{15162 - 1516 * C_{rate}}{RT_{bat}}\right)\right)^{\frac{1}{0.824}} * Q_{loss}^{\frac{0.824-1}{0.824}}(k) * \frac{T_s * I_{bat}(k)}{3600} \tag{29}$$

where the sampling time $T_s$ is 1 s, the price of battery capacity loss $C_{bat}$ is 694.4 USD / Ah, and the price of electric $C_{ele}$ is 0.1685 USD / kWh. To avoid the over-charge and over-discharging of batteries and supercapacitors, the operation windows of batteries and supercapacitors are constrained by 10% to 90% SOC of batteries and 50% to 99% SOC of supercapacitors. $W_{sc}$ and $W_{bat}$ weight the state constraints violations. These two weights are used to prioritize the prevention of over-charge and over-discharging than of battery degradation and electric cost. Thus, $W_{sc}$ and $W_{bat}$ are much higher than $C_{bat}$ and $C_{ele}$ to enable prioritization. $\sigma_{sc}$ and $\sigma_{bat}$ are the slack variables on the state constraints. When both constraints are met, and slack variables $\sigma_{sc}$ and $\sigma_{bat}$ are zero in the optimal solution. However, if constraints are violated, softened state constraints ensure the feasibility of the problem, and their associated linear weights can immediately penalize small constraint violations. $J(k)$ indicates the accumulative cost of the capacity loss of batteries and electric loss of resistances from time N to time k. According to Bellman's cost-to-go principle, by solving the DP backward, the minimal costs of all states along the entire driving cycle can be found and stored as a global optimal control strategy.

The reference power demands are predicted by collecting traffic information and predicting PLF on the cloud level with the vehicle dynamics model. As shown in Fig. 10, the discrete-time DP is applied to generate reference power profiles of batteries $P_{bat_{ref}}$ and supercapacitors $P_{sc_{ref}}$. The prediction horizon of the DP is set as 1200 seconds, and the DP is performed every 60 seconds with updated current states of the batteries and supercapacitors. The prediction horizon and update frequency are design parameters determined by the trade-off between computational cost and performance; With accurate models and predicted reference power demands, a longer prediction horizon and faster update frequency provide better performance but lead to a higher computational cost. The DP works as a block model predictive controller that generates optimal reference control signals with the 1200 seconds prediction horizon and allows the vehicle control unit at the vehicle level to follow the first 5% of the reference control signals to charge and discharge batteries and supercapacitors.

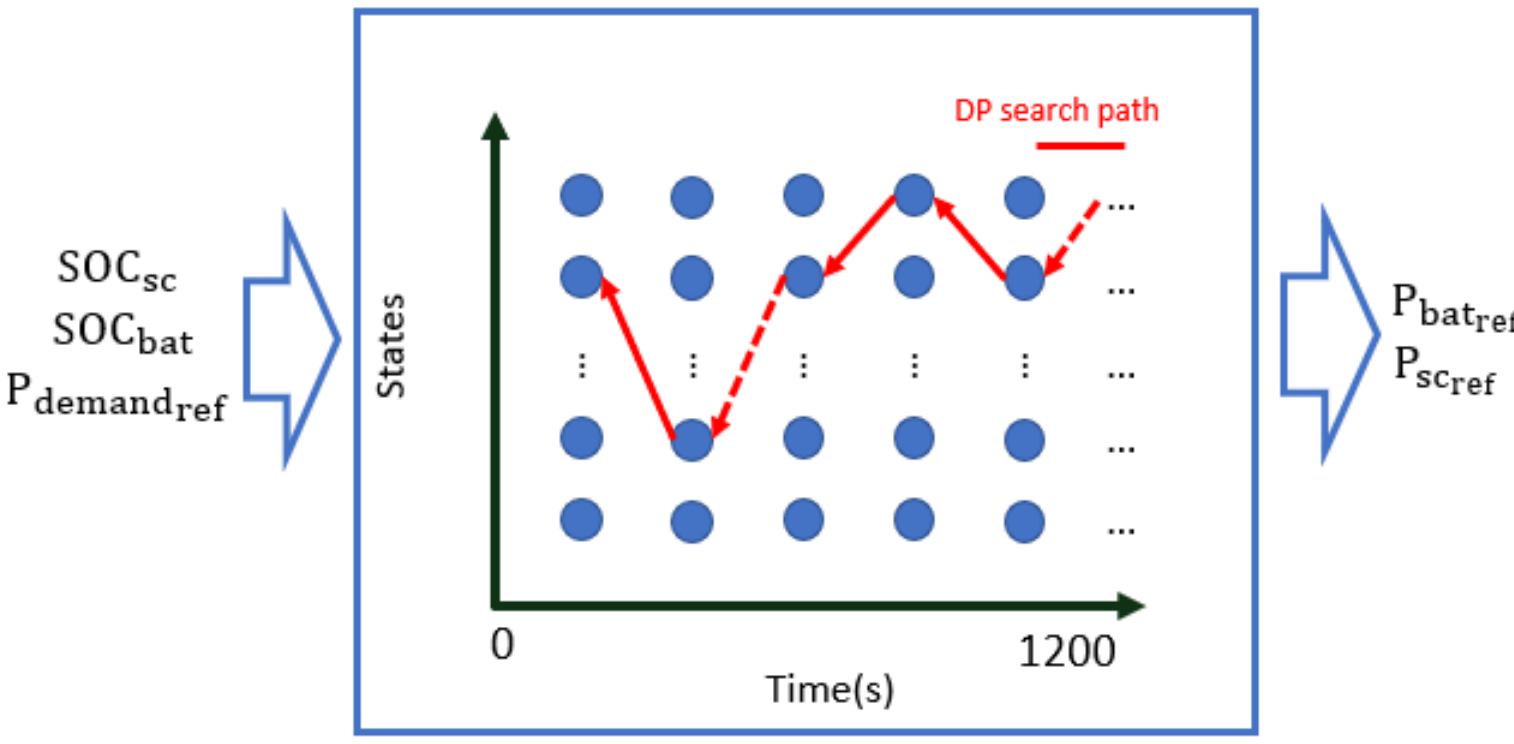

Fig. 10. The schematic diagram of the discrete-time DP.

3.2.2 Control Rule Extraction for DP Results

Due to the disturbances and the predictive errors of PLF and traffic information, the actual power demands may differ from the reference power demands used in cloud DP. Thus, on the vehicle side, a rule-based controller is used to handle the errors between the actual and reference power demands. The rule-based method is one of the most widely used energy management strategies to split power between batteries and supercapacitors under different conditions. When the demand powers are high, the supercapacitors are used to supply the powers to avoid the degradation of batteries. However, supercapacitors cannot keep a high-power supply for a long time because of their relatively low energy density. The supercapacitors need to be charged by regeneration power or batteries to keep the power supply ability for the future. Thus, the critical point of designing a rule-based controller is to figure out the charge and discharge conditions of the supercapacitors. The results of an offline DP are usually used to generate an optimal rule-based controller. By plotting one of the DP results in Fig. 11, we see that the relationship between the power demands and supply powers of supercapacitors is almost linear. Because the DP provides globally optimal results, an optimal rule of the rule-based controller can be obtained by fitting the relationship between the power demand and supply power of supercapacitors with the linear regression method.

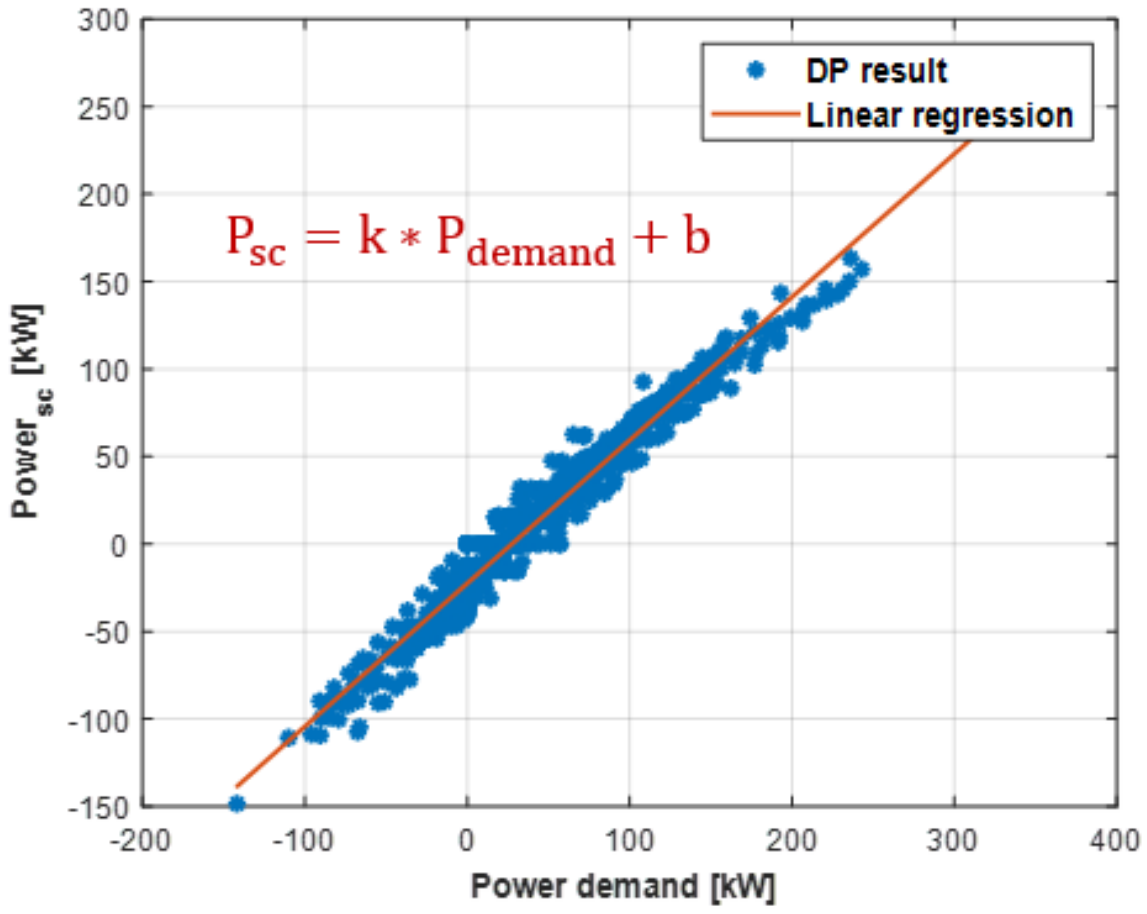

Fig. 11. DP result of $P_{sc}$ $vs.$ $P_{demand}$ with 100 % passenger load factor.

The difference between actual and reference power demand is represented as,

$$\Delta P_{demand} = P_{demand} - P_{demand_{ref}} \tag{30}$$

3.2.3 Rule-based Control at Vehicle Level

After the optimal rule, i.e., the linear relationship between power demand and supply power of supercapacitors, is extracted from the DP result, $\Delta P_{sc}$ and $\Delta P_{bat}$ are calculated in the rule-based controller to handle the $\Delta P_{demand}$ by following the flow chart presented in Fig. 12. The rule-based controller uses the $SOC_{sc}$ and the difference between reference and true power demands to determine its control strategy. To avoid the over-charge and over-discharge of the supercapacitors, 1% SOC buffer zones are set for the lower and higher bounds of the supercapacitor SOC operation range. The SOC operational range of supercapacitors is 51%-99%. When the $SOC_{sc}$ is smaller than the lower bound, only charging is allowed to be applied on the supercapacitors. Thus, if $\Delta P_{demand}$ is higher than zero, the batteries are used to handle the whole $\Delta P_{demand}$. Otherwise, the extracted rule from DP will be adopted. Vice versa, when the $SOC_{sc}$ reaches the upper bound, supercapacitors only discharge their energy. Under this condition, if $\Delta P_{demand}$ is smaller than zero, which means additional change energy is available, the batteries will handle the whole $\Delta P_{demand}$. Otherwise, $\Delta P_{demand}$ will be optimally split by the extracted control rule.

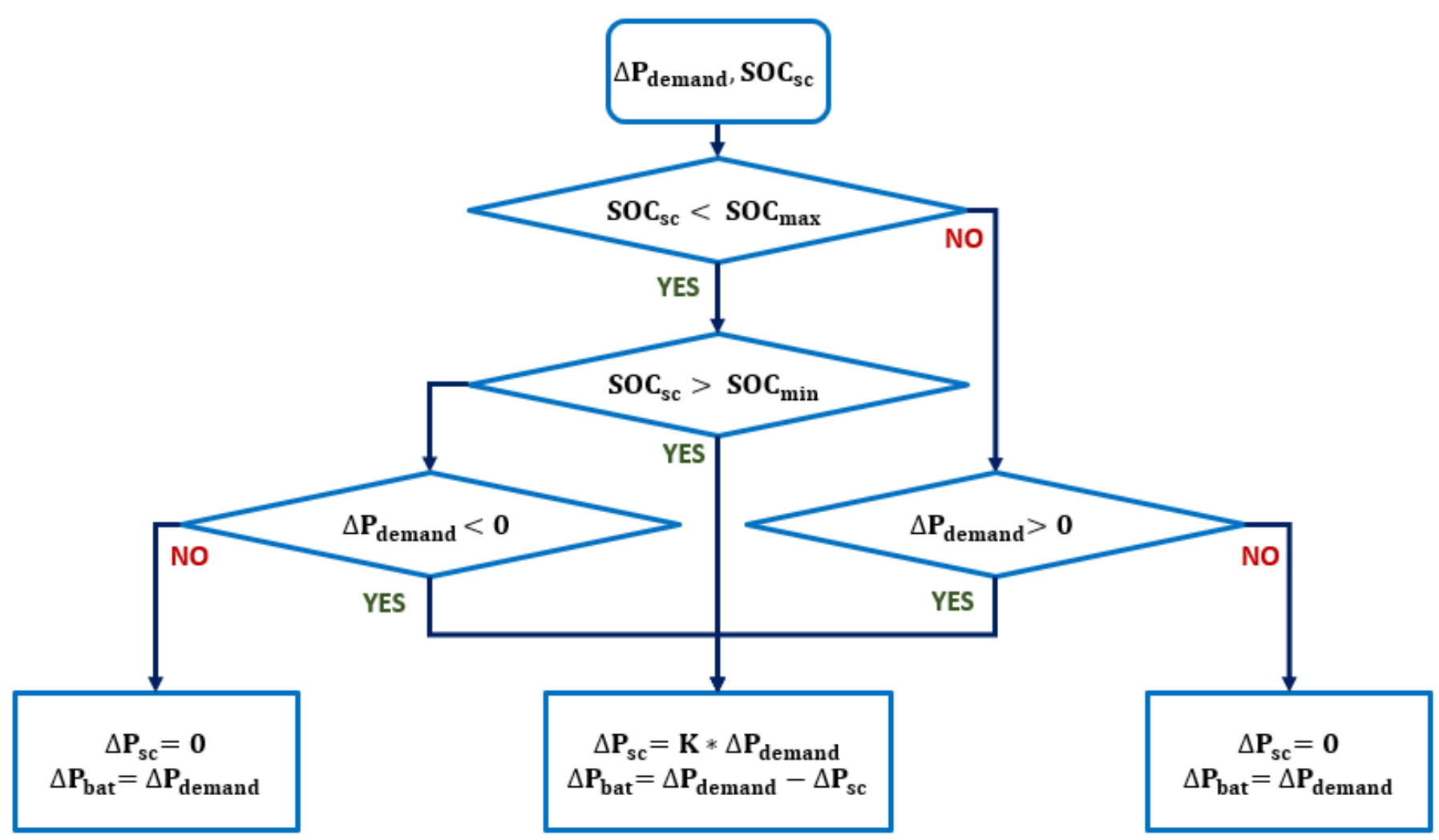

Fig. 12. The flow chart of the rule-based control extracted from DP results.

By combining the DP and rule-based control results, the optimal powers supplied from batteries $P_{bat}$ and supercapacitors $P_{sc}$ are updated as,

$$P_{sc} = P_{sc_{ref}} + \Delta P_{sc} \tag{31}$$

$$P_{bat} = P_{bat_{ref}} + \Delta P_{bat} \tag{32}$$

## 4. Results and Discussion

In this study, real-time energy management strategies are tested and validated by the typical China Bus Driving Cycle (CBDC) [12]. The CBDC speed profile is converted to different power demand profiles with different passenger load conditions. The CBDC speed profile and power profiles with different PLF are presented in Fig. 13. The CBDC speed profile shows frequent acceleration and braking events with the maximum speed below 15m/s (i.e., 53kg/h or 33mph). Power demand varies significantly along with different PLF as the 100% load factors results in high peaks at high speed and deep valleys at braking.

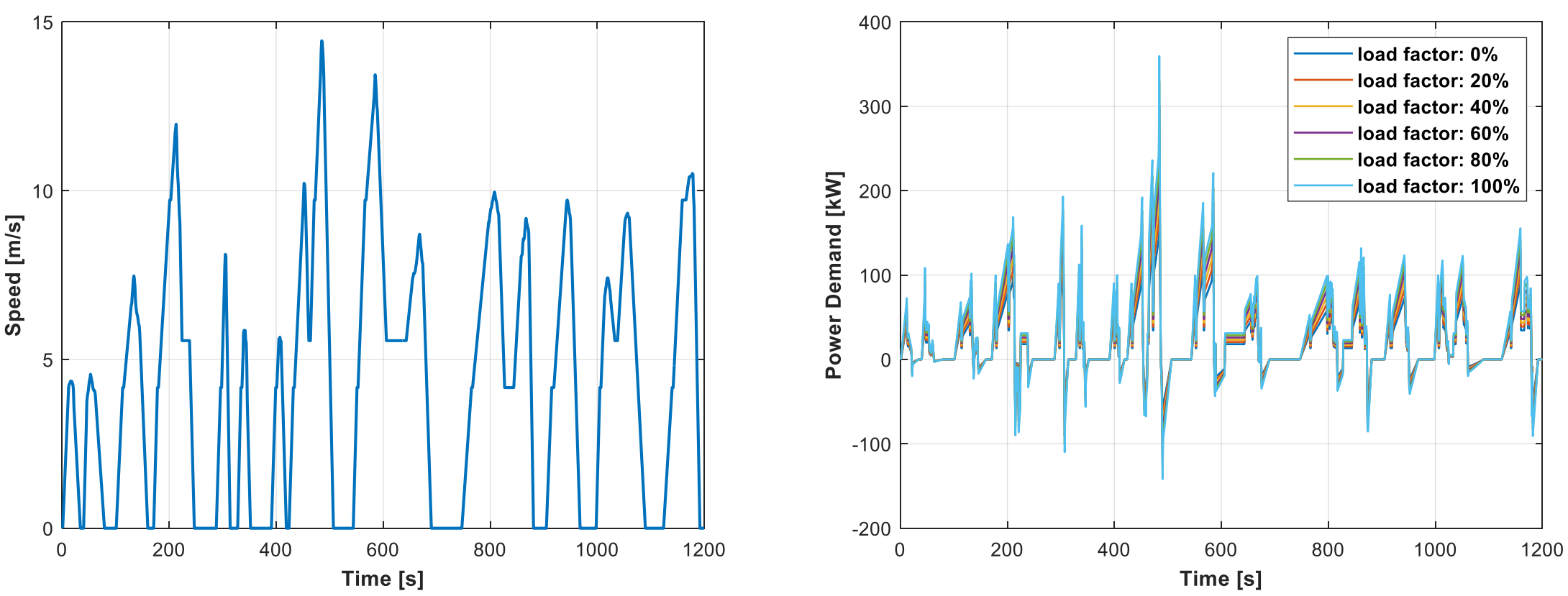

(a) CBDC speed profile

(b) CBDC power profile with different passenger load factors

Fig. 13. China Bus Driving Cycle.

### 4.1 Battery Aging Effect

In order to analyze the impact factors of an optimal energy management strategy generation of hybrid city buses, a comparison study is performed in this subsection. The DP approach is utilized as the energy management strategy in the battery aging effect discussion.

The battery degradation among the life cycle can be summarized into three phases: 1) SEI formulation; 2) SEI thickening; 3) Li-plating [29]. The battery degradation rate is high in phases 1) and 3). However, EV and HEV applications only consider phases 1) and 2) to satisfy the mileage requirement in general, because once batteries are degraded to 80% of the nominal capacity, the batteries will be replaced. As a result, only phases 1) and 2) are considered in this study. An aged battery, as in phase 2, has a lower degradation rate than a fresh battery, as in phase 1. This is also aligned with the battery degradation model, Eq. (7), used in this study and the simulation results shown in Fig. 14. Based on Eq. (7), the delta capacity loss in a sample time depends on the battery temperature, C-rate, and previous accumulated battery capacity loss. As shown in Fig. 14, with fixed battery temperature and C-rate, the delta capacity loss of a battery in a sampling time decreases as the battery ages (i.e., y-axis drops as state-of-health (SOH) decreases from 100% to 50%). SOH = 100% corresponds to a brand-new battery and SOH = 0 % means 20% capacity loss, as known as end-of-life (EOL). Namely, the cost of battery capacity loss in a sampling time will decrease progressively as the battery ages in EV/HEV applications. In addition, using batteries to charge supercapacitors leads to extra electrical costs because of the electrical energy losses due to the resistances of supercapacitors. Because a relatively aged battery can be used to handle a higher charge and discharge current than a brand-new battery, we expect that the battery aging effect impacts the formulation of an optimal energy management strategy. By involving a battery degradation model, one of the advantages of the proposed EMS is the ability to dynamically adjust its control strategy as batteries ages. If the SOH of the batteries is low, the cost function Eq. (16) used in the DP will put more effort into minimizing the electric loss rather than avoiding the battery degradation due to the high charge or discharge current.

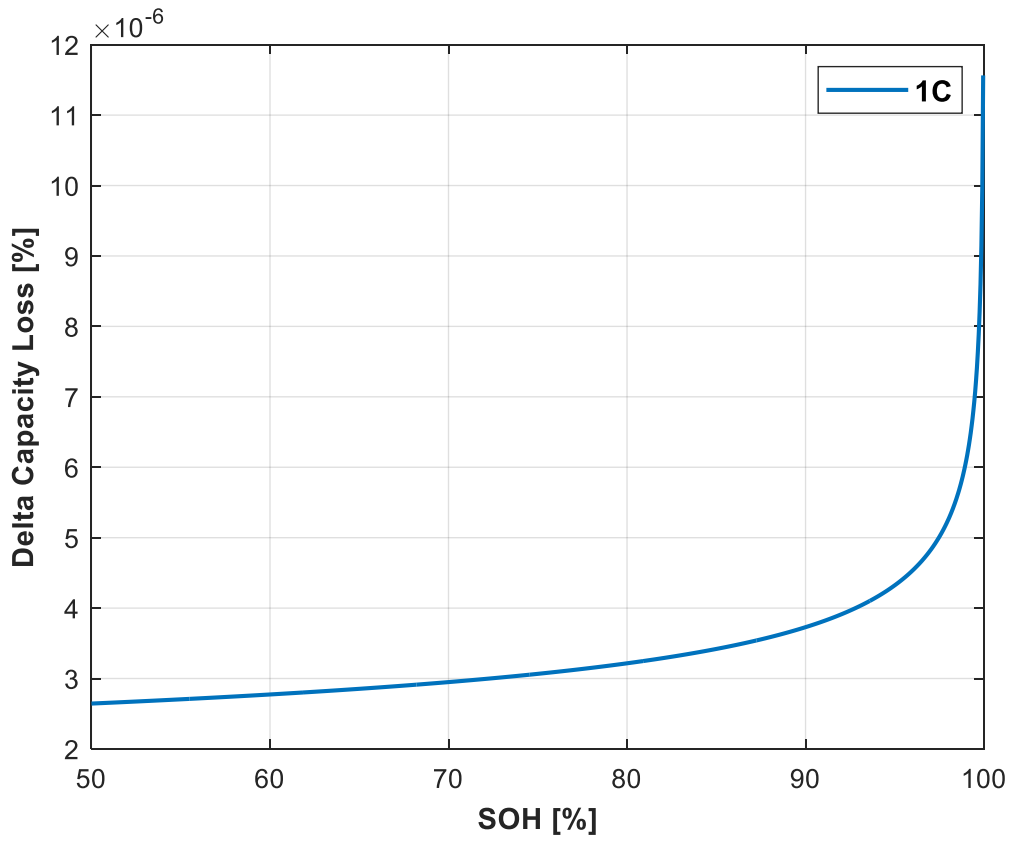

Fig. 14. The delta capacity loss of a battery with 1 C discharge current from 100 % SOH to 50 % SOH.

To further validate and analyze the aging effect of the batteries on the optimal energy management strategy generation, the control strategies that consider the aging effect are tested in this study. The dynamic programming generates two control strategies with the same power profiles but different battery SOH values. These two control strategies are then tested by the same vehicle operating condition. The results are shown in Fig. 15.

As shown in Fig. 15, the strategy (Strategy 1) represented by the blue line is generated with a 100 % PLF and 100 % SOH, and the strategy (Strategy 2) represented by the red line is generated with a 100 % passenger load factor and 0 % SOH. Both strategies are tested by an operation condition with a 100 % load factor and 100 % battery SOH. Because Strategy 1 is generated from a 100 % SOH, which has a higher battery degradation rate than the one from a low SOH condition, it puts more effort into saving battery degradation cost than Strategy 2 does. As shown in Fig. 15 (a), (b), (c), Strategy 1 tends to discharge the supercapacitors with higher power at high power demand ranges than Strategy 2 does to avoid discharging the battery with high C-rates. In Fig. 15 (e) and (c), at low power demand ranges, Strategy 1 tends to charge the supercapacitors and lets the SOC of the supercapacitors reach to values equal to or even higher than what Strategy 2 achieves. As shown in Fig. 15 (f), the total costs of Strategy 1 and strategy 2 are 3.8573 USD and 3.8795 USD, respectively. The result indicates that the strategy with a correct SOH value only brings 0.58 % of improvement than the strategy generated with a wrong SOH value dose. Namely, in this test, the aging effect of a battery does not play an important role in the optimal control strategy generation. The result is surprising as we would expect larger improvement with correct SOH estimation.

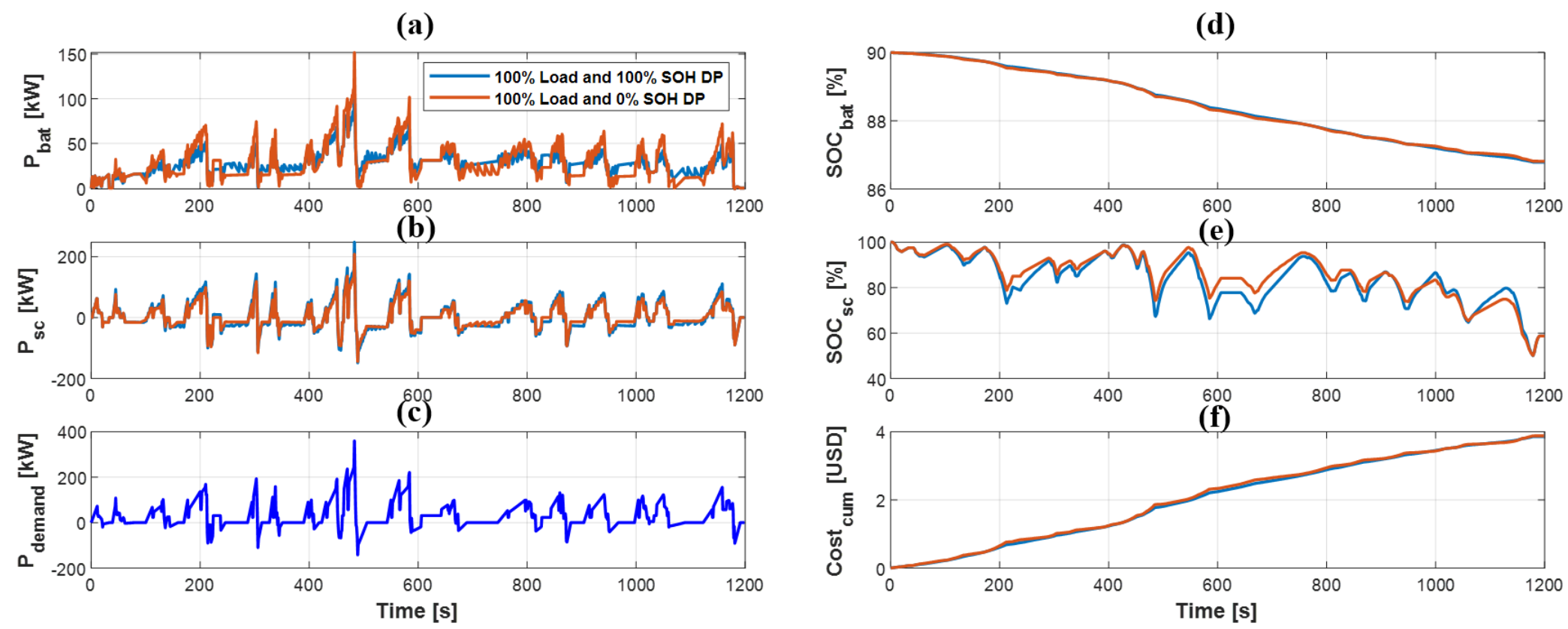

Fig. 15. Aging effect comparisons of 100% SOH and 0% SOH DP strategies by using a 100% SOH operation condition: (a) power profiles of batteries, (b) power profiles of supercapacitors, (c) power demand profile, (d) SOC profiles of batteries, (e) SOC profiles of supercapacitors, and (f) total cumulated cost.

To further investigate the logic behind the result, the costs along the CBDC using the same control strategy with 100% and 0% SOH are plotted and shown in Fig. 16. As shown in Fig. 16, the real-time cost of capacity loss with 0% SOH is much less than the cost with 100% SOH. The result matches the expectation. However, although the cost of battery capacity loss drops a lot from 100% SOH to 0% SOH, the real-time cost is still much higher than electrical loss. Namely, during the optimization, more efforts are put on minimizing the cost of the capacity loss rather than the electrical loss due to the resistance in both cases. As a result, similar control strategies are gained even they are generated from different SOH conditions. In conclusion, the SOH of a battery will not significantly impact the optimal energy split strategy generation in the studied vehicle unless the manufacture cost of the Li-ion batteries substantially decreases or electricity price increases in the future.

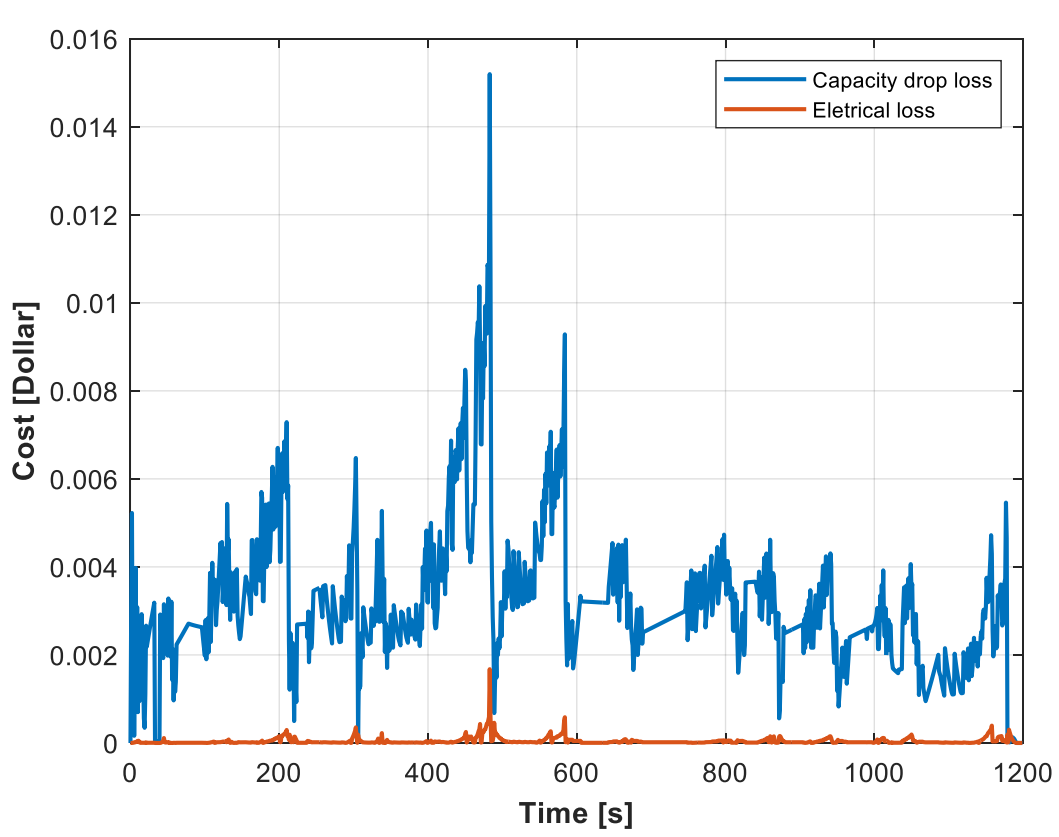

(a) Cost with 100 % SOH Batteries

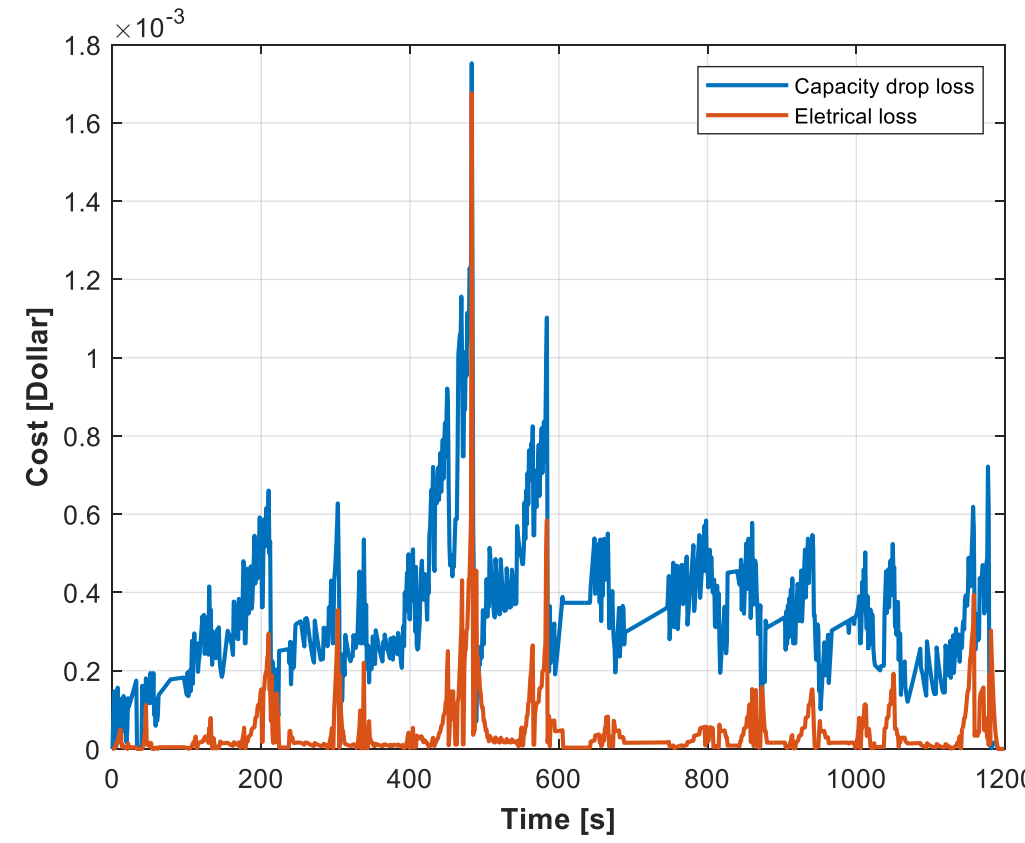

(b) Cost with 0 % SOH Batteries

Fig. 16. Real-time cost comparison with different SOH of Batteries.

4.2 Passenger Loading Effect

As shown in Fig. 13 (b), a higher load factor leads to higher power demands. Simultaneously, the higher power demands require higher C-rates that batteries and supercapacitors have to supply. As shown in Fig. 17, the battery capacity loss cost increases exponentially as the C-rate increases. Thus, as the PLF increases, more control efforts should be put on to avoid charging or discharging batteries with high C-rates to minimize the battery capacity loss.

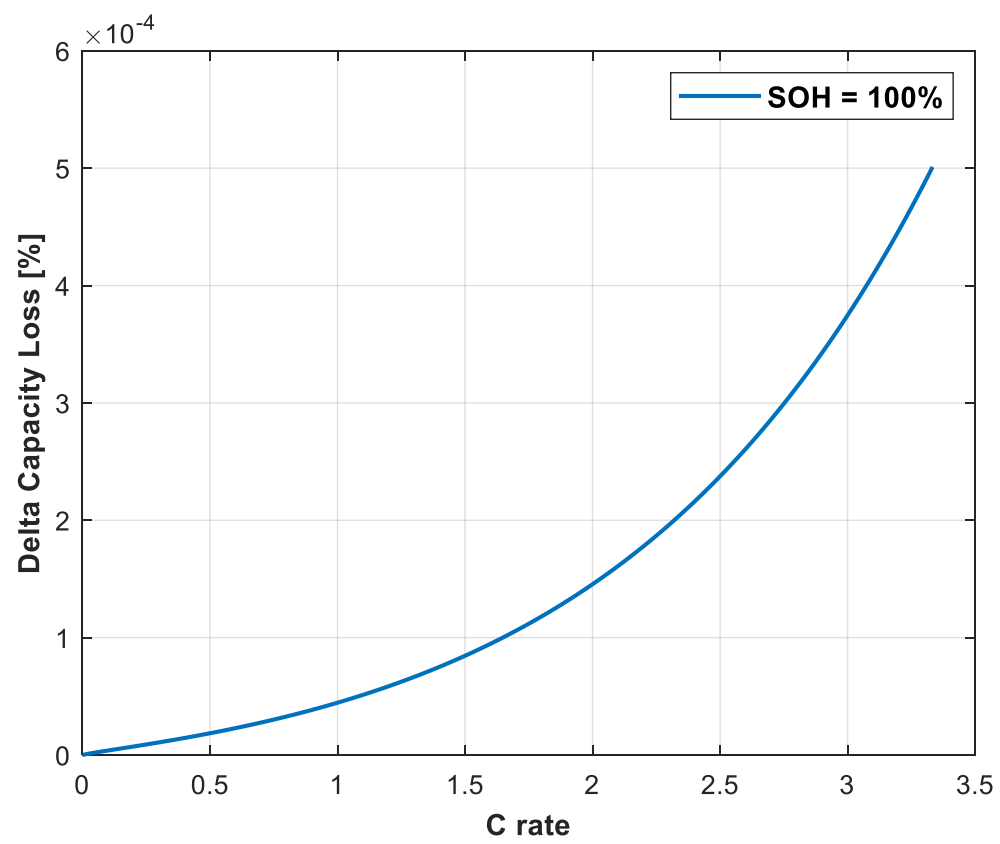

Fig. 17. The delta capacity loss of a battery with 1 C to 3.5 C discharge current and 100 % SOH.

Two control strategies are generated based on 100 % load (Strategy 1) and 0 % load (Strategy 2), respectively, with the dynamic programming algorithm to investigate the effect of the PLF. These two control strategies are then tested with a 100 % load and 100 % battery SOH condition. As shown in Fig. 18, there is a large difference between the two strategies. In Fig. 18 (b) and (c), when power demands are low, Strategy 1 is more likely to charge or save the energy of supercapacitors than Strategy 2 does. In Fig. 18 (c) and (e), by keeping high SOC values of supercapacitors during low power demand range, Strategy 1 enables supercapacitors to provide more power supply than batteries when power demands are high. As a result, in Fig. 18 (f), the total costs of Strategy 1 and strategy 2 are 3.8573 USD and 4.1369 USD, respectively. Without considering the loading effect, Strategy 2 introduces 7.24 % more cost than Strategy 1 does. This large percentage difference caused by the different passenger loading effect indicates the importance of the loading effect. It shows that the PLF should be considered while designing an optimal energy control strategy.

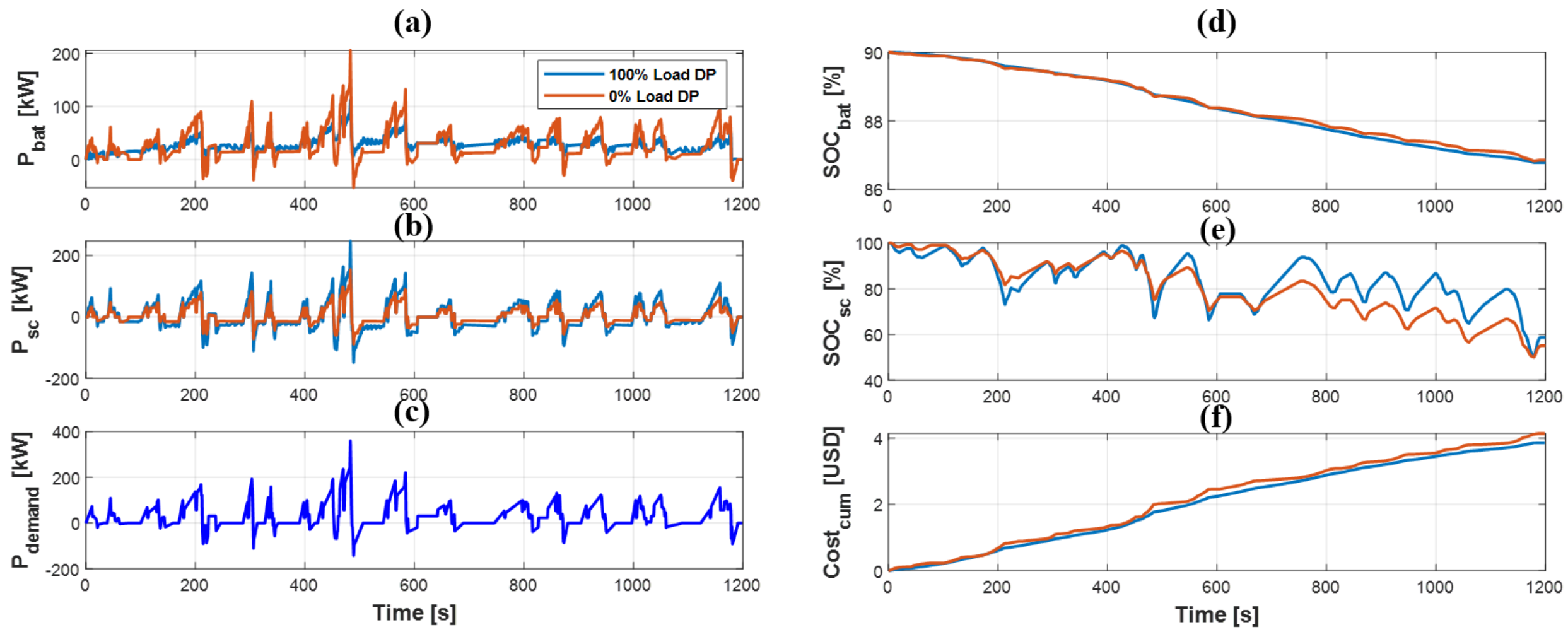

Fig. 18. Loading effect comparisons of 100 % load and 0% load DP strategies basing on CBDC and 100 % load factor: (a) power profiles of batteries, (b) power profiles of supercapacitors, (c) power demand profile, (d) SOC profiles of batteries, (e) SOC profiles of supercapacitors, and (f) total cumulated cost.

Besides, by extracting rules from DP, the fitted functions with CBDC and different PLF are shown in Table 4 to demonstrate the impacts of PLF on the rule update of rule-based controllers. The function in each row is fitted using the DP results considering the specific PLF from that row.

Table 4

The rule of rule-based controller with different passenger load factors.

| Passenger Load Factor (%) | Fitted Function (W) |
|---|---|
| 0 | $P_{sc} = 0.8391 * P_{demand} - 11803$ |
| 20 | $P_{sc} = 0.8277 * P_{demand} - 13808$ |
| 40 | $P_{sc} = 0.8270 * P_{demand} - 15948$ |
| 60 | $P_{sc} = 0.8268 * P_{demand} - 18247$ |
| 80 | $P_{sc} = 0.8240 * P_{demand} - 20402$ |
| 100 | $P_{sc} = 0.8183 * P_{demand} - 22476$ |

As shown in Table 4, with a higher PLF, the slop and intercept value of a fitted function are smaller. Because a higher passenger load factor leads to higher power demands, supercapacitors need to supply more power at the high-power demand ranges to minimize the capacity loss of batteries. The optimization strategies tend to charge supercapacitors with more powers or save the energy of supercapacitors when power demands are low. Thus, for the proposed EMS, with the updated rule of the rule-based controller at different passenger load conditions, the error between predicted and true power demands can be handled in an optimal way.

## 4.3 Passenger Load Factor Prediction

The average and regression models are trained using the passenger load data of Guangdong Line 11 city buses from 08/01/2014 to 12/14/2014. In the subsection, these models are tested by one-week passenger load data from 12/15/2014 to 12/22/2014. From the test, the comparisons between actual and predicted PLF of each model are presented in Fig. 19.

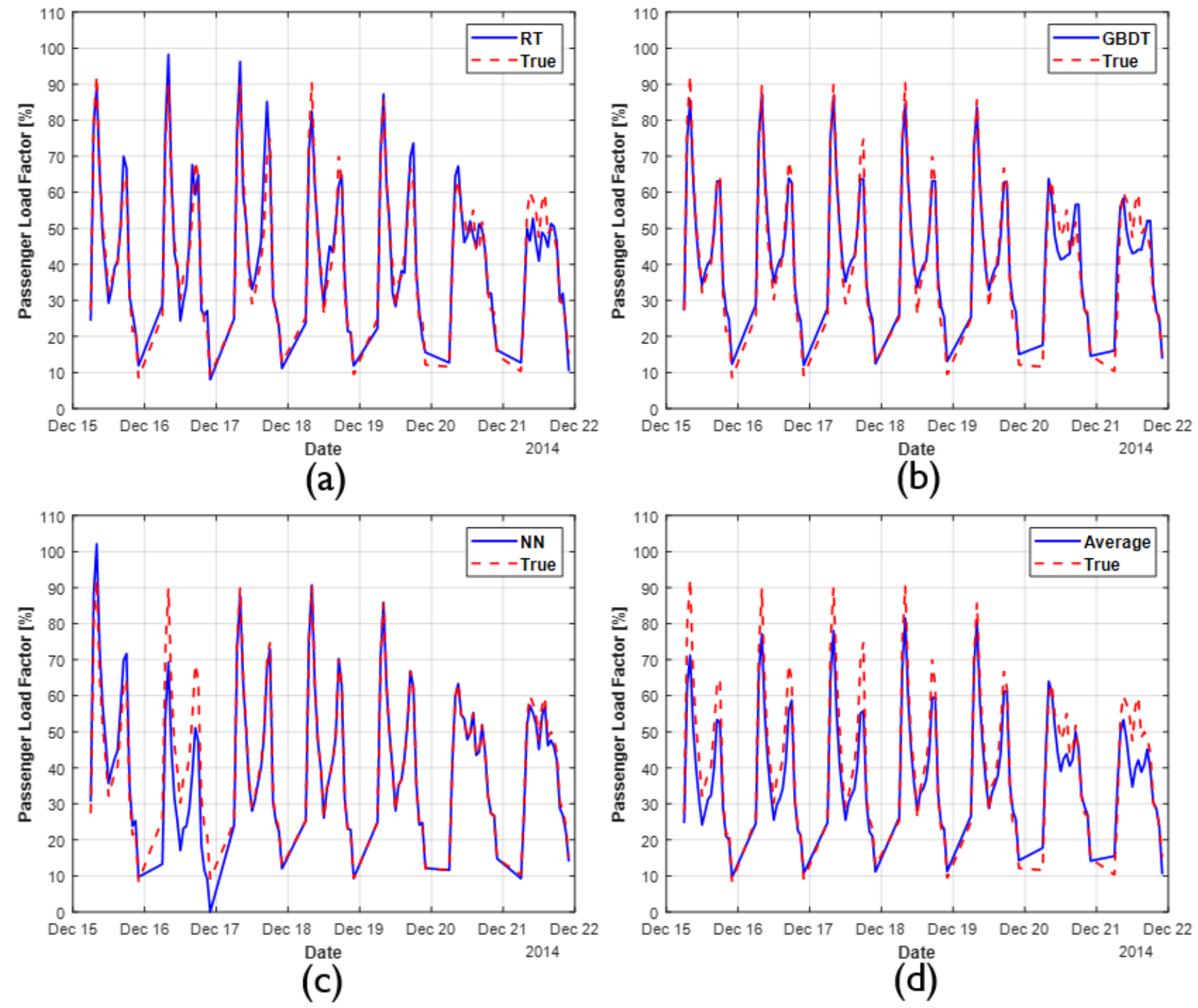

Fig. 19. Passenger load factor prediction test for RT, GBDT, NN, and Average models with one-week Guangdong line 11 city bus passenger load data from 12/15/2014 to 12/22/2014.

As shown in Fig. 19 (a) and (b), RT and GBDT prediction results have good alignment with the true results. In Fig. 19 (c), NN model has extremely good prediction accuracy on all the days except Dec 16 and 17. In Fig. 19 (d), the average model underestimates nearly all the peak load factors.

To further analyze the accuracies of prediction results, the accuracy of the forecast is evaluated by the root mean squared error (RMSE),

$$\mathrm{RMSE} = \sqrt{\frac{1}{m}\sum_{i=1}^{m}(y_i - \hat{y}_i)^2} \tag{33}$$

where y and $\hat{y}$ are the true and predicted PLF, and m represents the number of the predictions. RMSE of each day and the whole week for each model are calculated and presented in Table 5. Besides, the variances of these RMSE of each model in the whole week are gained by using,

$$V = \frac{1}{N-1}\sum_{i=1}^{N}\left|\mathrm{RMSE}_i - \overline{\mathrm{RMSE}}\right|^2 \tag{34}$$

where N is the number of days in a week, and $\overline{\mathrm{RMSE}}$ is the mean RMSE of each model.

Table 5

The RMSE of each day and the whole week for each model.

| RSME<br>Model | Mon | Tu | Wed | Th | Fri | Sat | Sun | Total | Variance |
|---|---|---|---|---|---|---|---|---|---|
| RT | 3.0113 | 5.7436 | 6.6337 | 4.2158 | 4.2971 | 3.8723 | 6.0119 | 4.9776 | 1.7285 |
| GBDT | 3.2303 | 3.2929 | 4.2915 | 3.6366 | 2.7078 | 6.5581 | 7.6120 | 4.7988 | 3.4977 |
| NN | 5.4844 | 14.911 | 1.3670 | 0.0665 | 0.0091 | 0.0017 | 2.1439 | 6.0816 | 29.454 |
| Average | 10.313 | 7.1619 | 8.5135 | 4.7929 | 3.7301 | 5.1463 | 9.1172 | 7.3353 | 6.1290 |

As shown in Table 5, the three regression models all work very well. NN model provides very accurate predictions from Wednesday to Sunday. The performances of the RT model are good and very stable, with the lowest RSME variance. The prediction results of GBDT have the lowest RMSE for the whole week, and the variance is the second-best only behind the RT model. As a result, the GBDT model is selected as the most suitable model for the PLF prediction in this study considering the benefits the GBDT model brings, including, 1) high prediction accuracy, 2) fast prediction speed with only modest memory requirement, 3) no requirement on the performance of complex feature engineering, and 4) ability to handle a mixture of feature types.

Comparing with the city bus passenger loading factor prediction models in the existing literature, such as [19], [20], the proposed three regression models have better prediction performance. Instead of only using the historic passenger loading data to train models, the proposed models consider more factors, including day-of-week, time-of-day, weather, temperatures, wind levels, and holiday information. Besides, the proposed models have the capability to model any complex non-linear functions.

4.4 Learning Curves and Feature Importance Ranking

In the machine learning domain, more data usually leads to a better prediction result. One of the most famous quotes describing the power of data is that "We don't have better algorithms. We just have more data," which Google's Research Director Peter Norvig claimed. However, data is not always available. Thus, this subsection focuses on discussing how many data samples are necessary and what features are important to accurately train PLF prediction model.

As shown in Fig. 20, a learning curve of selected GBDT model plots the PLF prediction errors of the training set against the errors evaluated on the validate data set with the same features. When the training set size is smaller than 750, the cross-validation errors decline rapidly as the training set size increase. It means the learning algorithm is suffering from high variance. In this case, getting more training data is very helpful. A training set with 750 samples is a critical point in this study, as the error percentage jumps to an acceptable value smaller than 10 %. As the training set size keeps increasing, the training errors and cross-validation errors tend to converge and become closer to each other. Getting more correlative features is likely to further decrease both training and cross-validation errors.

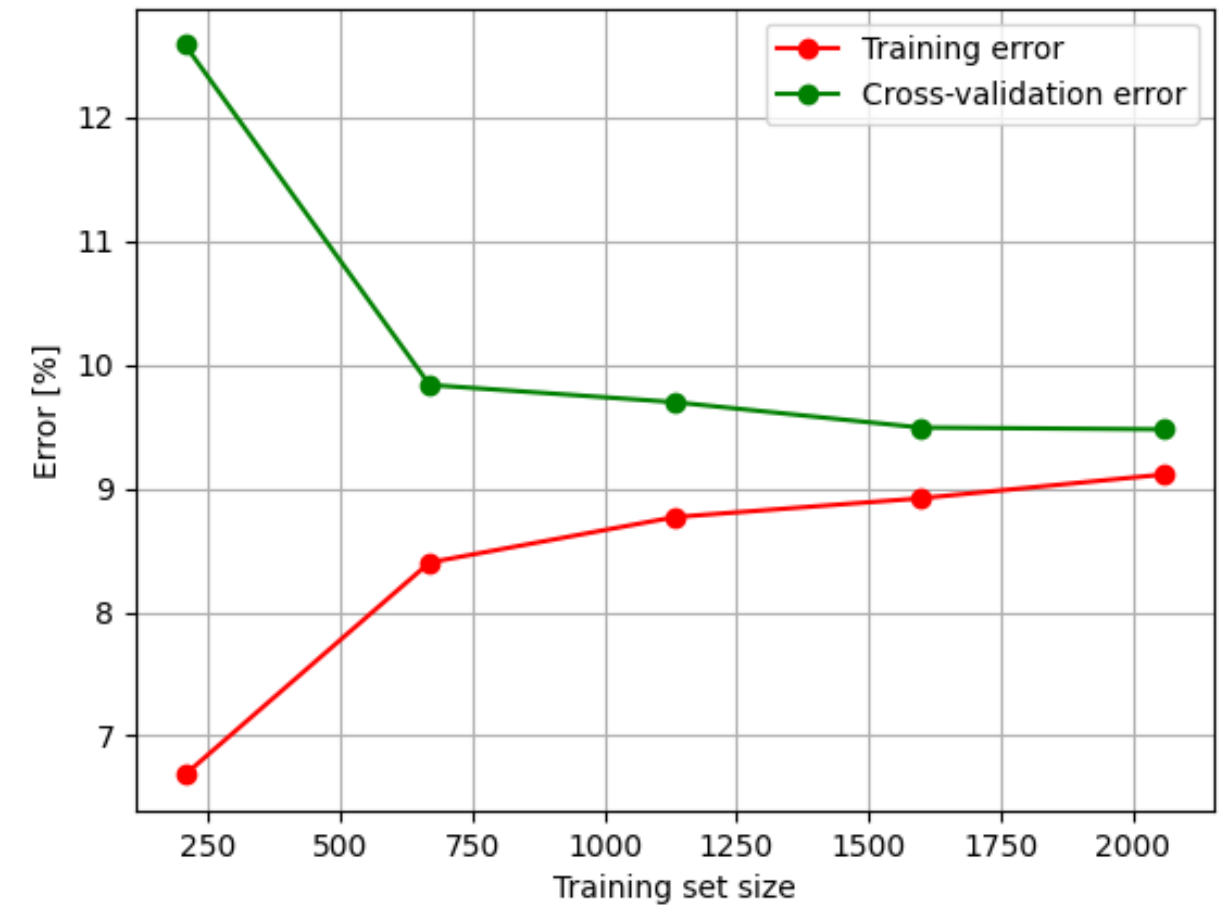

Fig. 20. Learning Curves of GBDT model.

Feature importance ranking reveals the contribution of each feature to the prediction accuracy. The selected GBDT algorithm is used to rank the feature importance of the PLF prediction. The ranking results are presented in Fig. 21 (a). The top three features are time-of-day, holiday information, and day-of-week. The holiday information is more important than the feature of day-of-week is because Saturday and Sunday are also considered as holidays in this study. As shown in Fig. 21 (b), these top three features contribute more than 90% of feature importance. The result is expected because Fig. 5 has already revealed the strong correlations among the PLF, time-of-day and day-of-week. Besides, the feature importance results provide quantified numbers to explain why the simple average model only taking time-of-day and day-of-week as inputs can provide reasonable prediction results. The importance of temperature, weather, and wind level shows that the commute of people taking public transport is less likely impacted by nature conditions. Thus, a reasonable PLF prediction model can still be developed, even if there is no temperature, weather, and wind level data because of data accessibility issues.

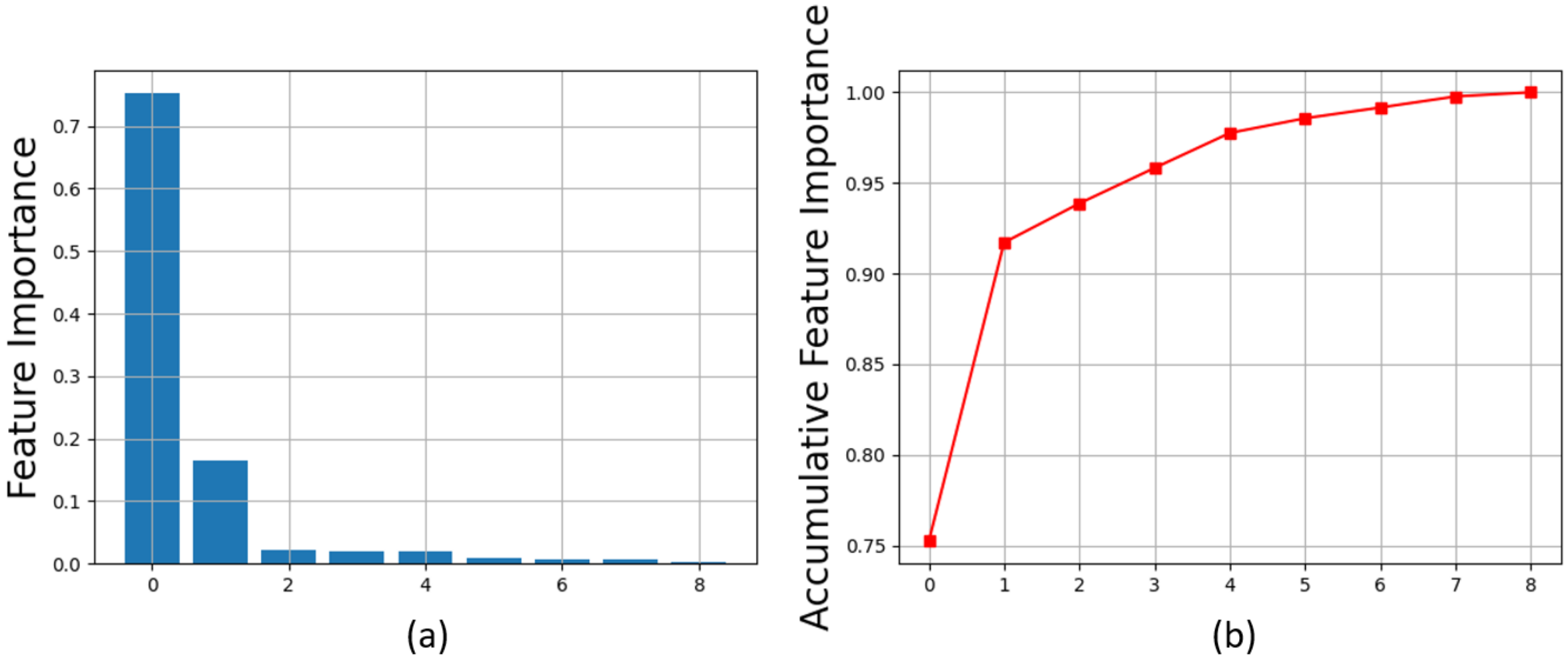

Fig. 21. Feature importance ranking and accumulative importance. (Features in order: 0. Time-of-day, 1. Holiday information, 2. Day-of-week, 3. The lowest temperature of the day, 4. The highest temperature of the day, 5. Average weather, 6. Daytime weather, 7. Weather at night, 8. Wind level.)

### 4.5 Comparison of Three Energy Management Strategies

In this subsection, the proposed method is evaluated and discussed with true passenger load data of Guangdong Transit Bus Line 11 on 12/15/2014 and CBDC. The real passenger load data at peak hour 8:00 AM and off-peak hour 12:00 PM are used in the test because the tests in these hours represent the most and least impacts of loading effect, respectively. The plot of true and GBDT predicted PLF on this day are presented in Fig. 22 (a). As shown in Fig. 22 (b), the predicted and true PLF aligns well over the 16 hours span without large error. The comparison results are summarized in Table 6. During the peak hour (i.e., 8:00 AM), the prediction error is under 10%. At noon, the error is 20%, which is large due to the low absolute load factor of around 30%.

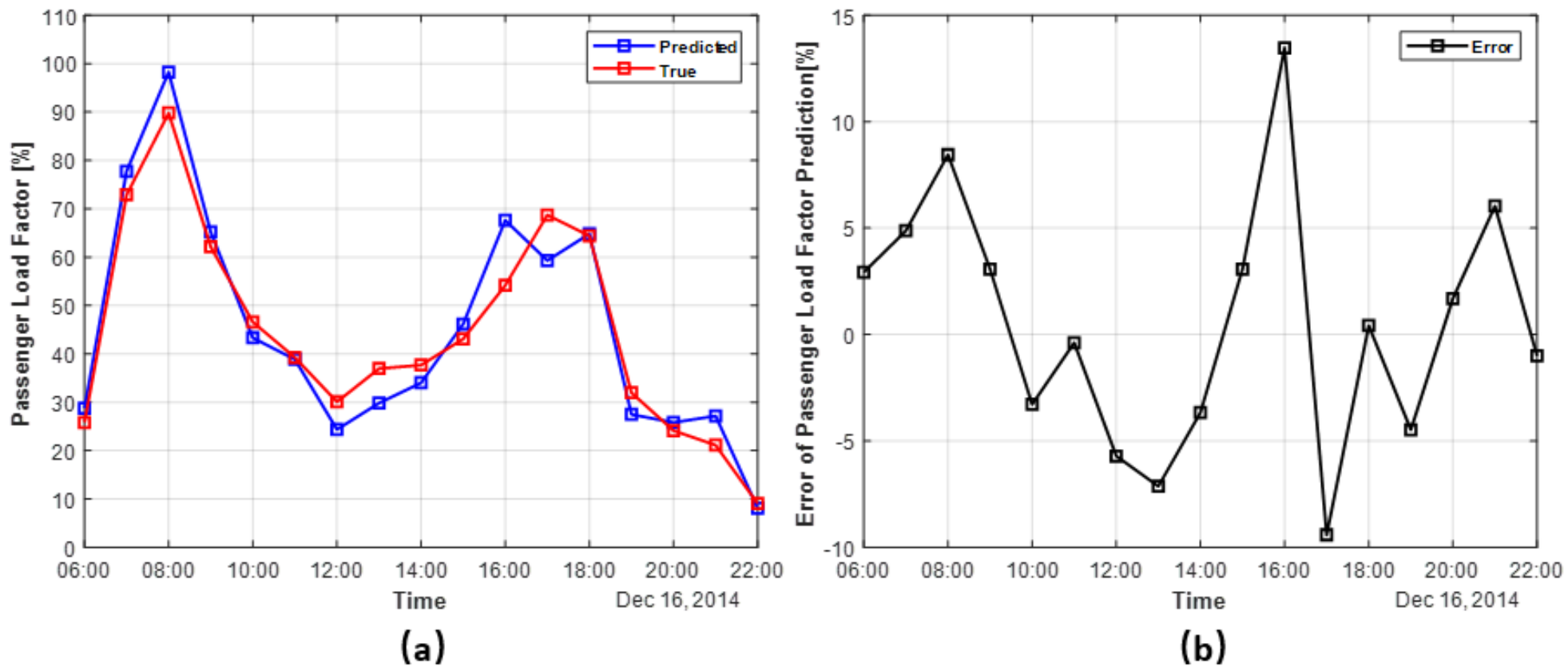

Fig. 22. Passenger load factors of Guangdong city bus line 11 on 12/15/2014: (a) predicted and true passenger load factors, (b) prediction error of passenger load factors with the proposed algorithm.

Table 6

Predicted and true passenger load factor, and prediction error at peak and off-peak hours of Guangdong city bus line 11 on 12/15/2014.

| Hour | Predicted Load Factor | True Load Factor | Absolute error | Error percentage |
|---|---|---|---|---|
| 8:00 AM | 98% | 90% | 8% | 8.9% |
| 12:00 PM | 24% | 30% | 6% | 20% |

The vehicle, battery, and supercapacitor models mentioned in section 2 are used in the validation. These models are widely used in literature to validate the designs of different EMS. The same models are used in [3] [9] [12] [21] to validate EMS for the hybrid city bus. Similar models are used to validate the designs of different EMS used in hybrid energy storage systems for electric vehicles [30] [31] [32]. Because EMS development is a system-level design, we believe models used in this study can well demonstrate and evaluate the performance and design of the proposed EMS.

In order to evaluate the performances of the proposed method, its test results are compared with the results of optimal DP strategies and the pure rule-based method. The structures of the two EMS are shown in Fig. 23. As shown in Fig. 23 (a), the first EMS is an offline dynamic programming strategy. The DP method is adopted in this paper to understand the minimum cost that a global optimization can bring to the entire driving cycle. All the future passenger load information is prior knowledge for dynamic programming. The optimal DP strategy takes true power demand profiles as the input which are generated from the actual PLF and speed profile. It is an ideal case and represent the best results that can be gained at these vehicle operating conditions. However, the lack of the methods of collecting highly accurate future PLF and traffic information impedes the implementation of the optimal DP strategy in practical applications. If the real driving conditions differ from the designed one, then the performance of a DP strategy cannot be guaranteed. Besides, the number of iterations to run a DP strategy is proportional to the product of the prediction horizon, size of the state space, and size of the action space. Due to the high computational cost of dynamic programming, the optimal DP strategy is only able to be implemented in offline. As shown in Fig. 23 (b), the second EMS is a pure rule-based method which addresses the weakness of the high computational cost by extracting the rules from the offline dynamic programming results and implementing the rules in an online rule-based controller. This rule-based method is proposed by [21], which demonstrates its good performance by comparing it with MPC controller, filtration based controller [33], and fuzzy logic controller [34]. The rule-based controller takes power demand and the SOC of supercapacitors as the inputs and splits the supply powers of batteries and supercapacitors during the operation. The rule-based controller used in the test is not benefited from passenger load prediction, but generated from the CBDC speed profile. The rule of the rule-based controller is fixed once it is developed and does not adapt to real-time varying conditions. Under other driving cycles that are different from the CBDC, the performance of the pure rule-based method may lead to significant downgrades. Thus, the rule-based method results in the test also an ideal case and represent the best performances that a conventional rule-based method can bring for an electric city bus.

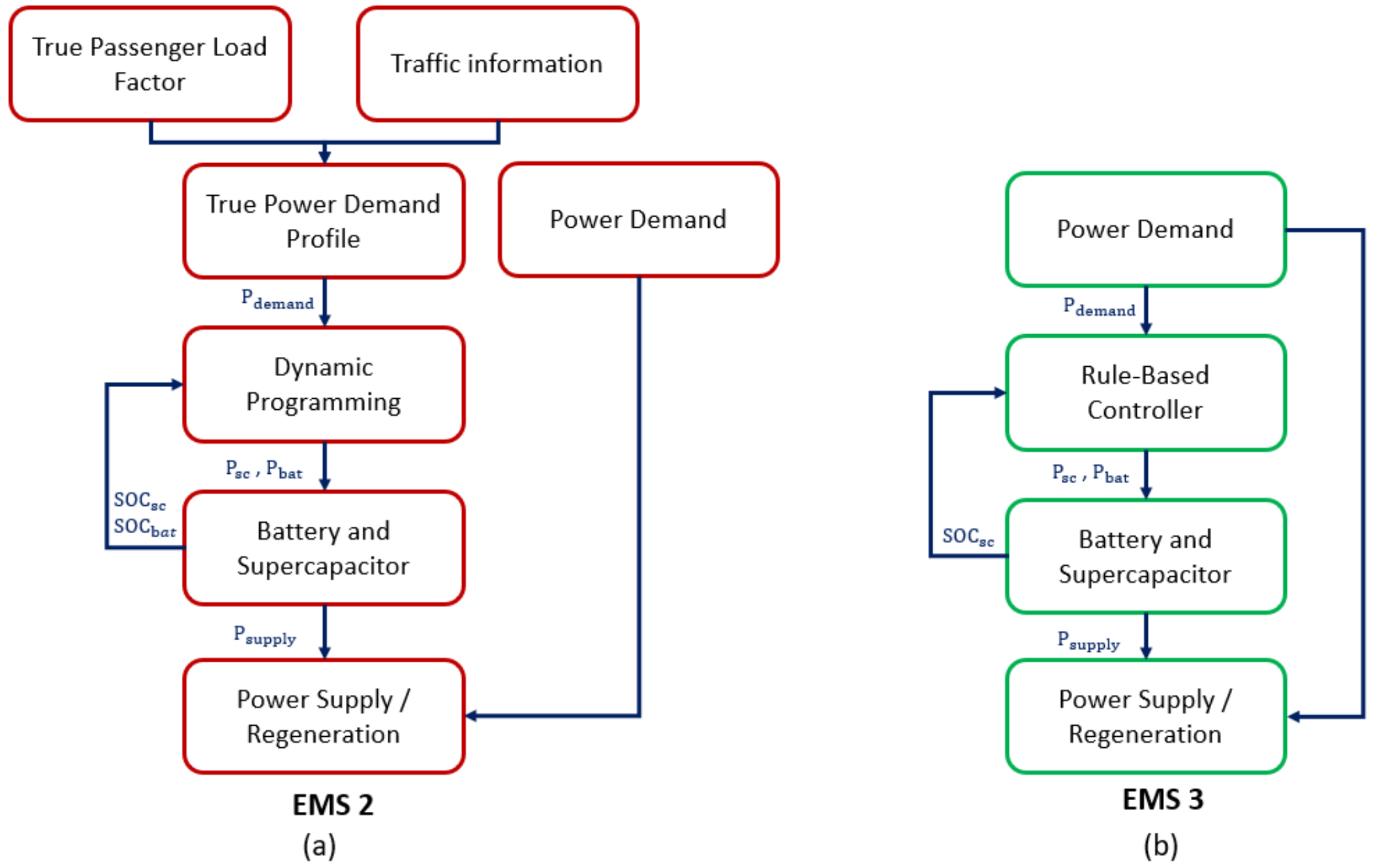

Fig. 23. Two EMS used in the comparison study as benchmarks: (a) DP with accurate power demand profile from the true passenger load information, (b) rule-based controller with DP rule-extraction and no passenger load prediction.

The comparison results at 8:00 AM are shown in Fig. 24 (f) and Table 7. These total costs of the proposed method, pure rule-based method, and optimal DP strategy are 3.6890, 4.1181, and 3.6885 USD, respectively. As shown in the supercapacitor SOC subplot, Fig. 24 (d) and (e), the proposed method provides a similar control strategy as the optimal DP strategy dose, while the pure rule-based method has a different SOC curve. Multiple supercapacitor SOC saturations are spotted between 400s and 1200s, which disable the possibility of discharge assistant to the battery at high power demand situations. This leads to the peaks from the pure rule-based method in the battery power subplot, Fig. 24 (a), during which the battery power sits at low levels for the other two methods. The proposed method and optimal DP method have a smoother battery SOC curve than the pure rule-based method. Even though the pure rule-based method shows the least cumulative cost at 450s, its supercapacitor SOC saturates at that time, and the cost immediately goes up, finally surpasses the costs from the other two methods. For the proposed method, even if there is an 8.9% prediction error of the PLF, only 0.0136% of the difference between the total costs of the proposed method and the optimal DP strategy is introduced. On the other side, the proposed method brings 10.4199% of cost reduction than the pure rule-based method dose in the peak-hour.

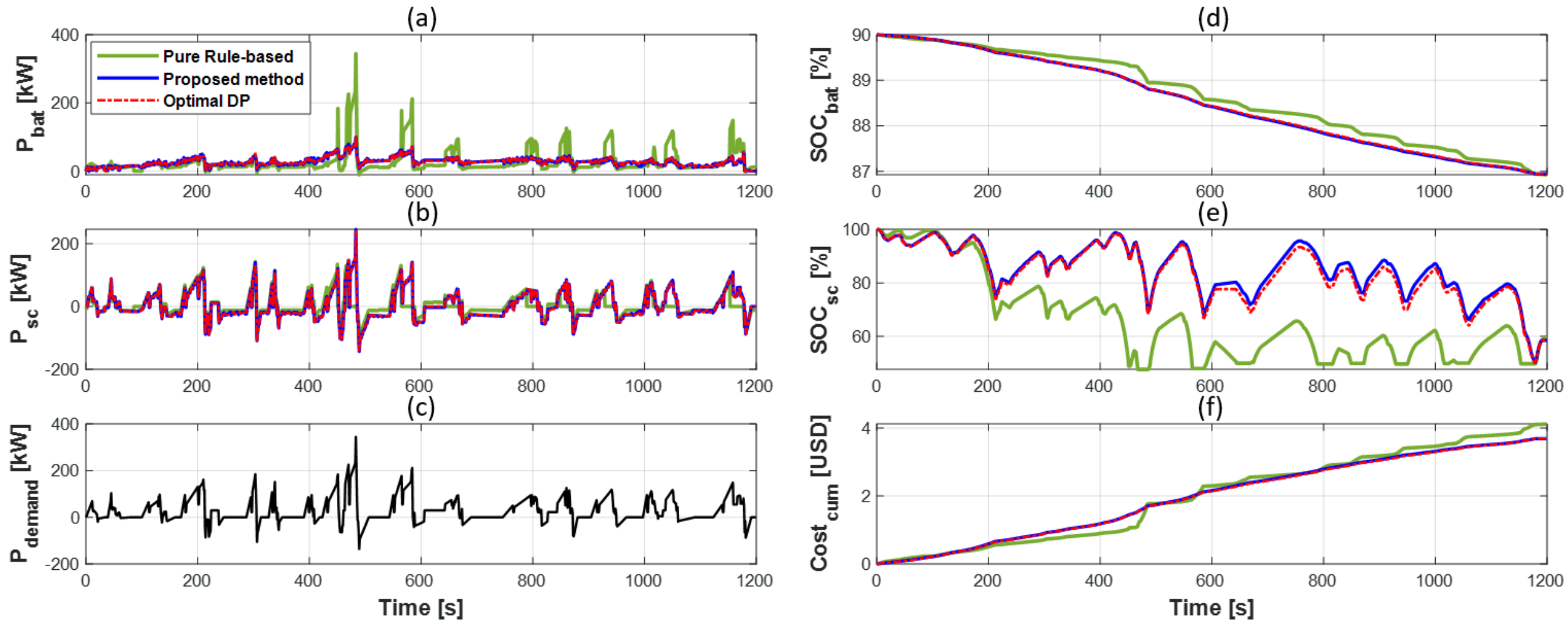

Fig. 24. Comparison of the proposed method, pure rule-based method, and optimal DP strategy with CBDC and true passenger load factor at 8:00 AM of Guangdong Transit Bus Line 11 on 12/15/2014: (a) power profiles of batteries, (b) power profiles of supercapacitors, (c) power demand profile, (d) SOC profiles of batteries, (e) SOC profiles of supercapacitors, and (f) total cumulated cost.

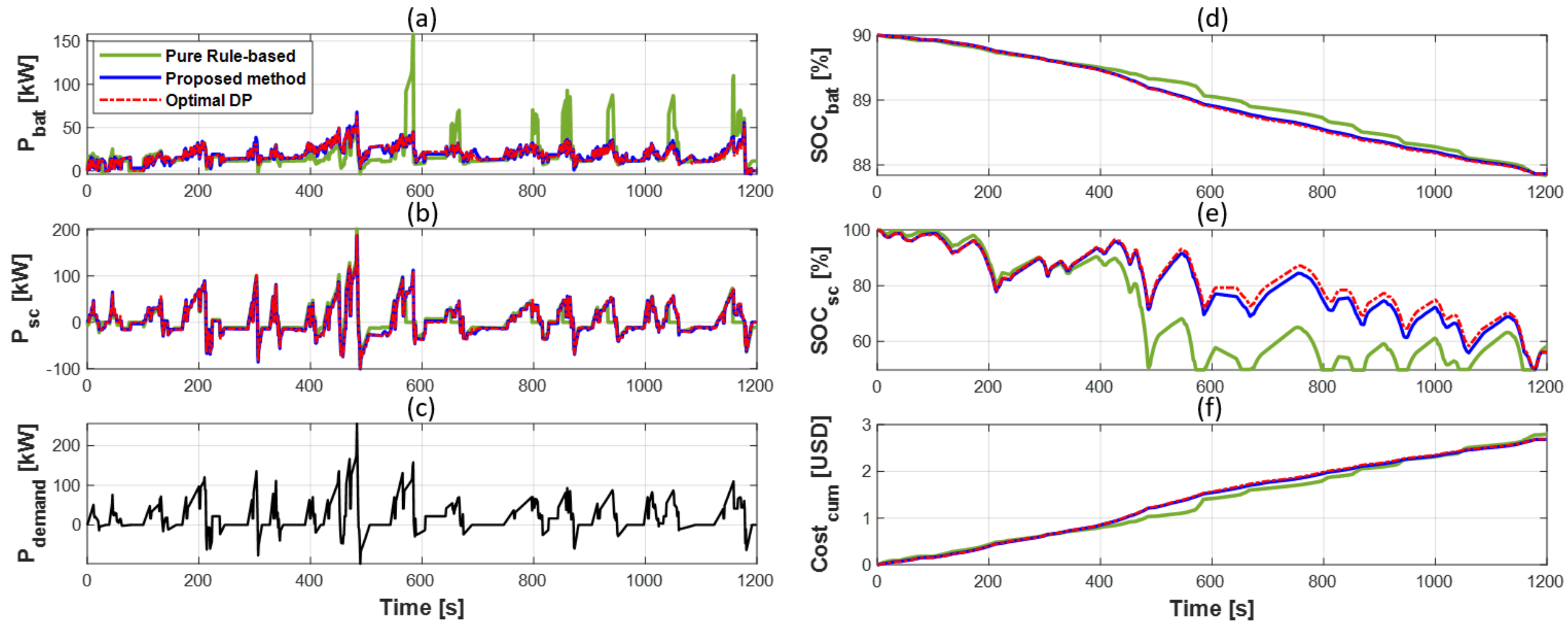

Fig. 25. Comparison of the proposed method, pure rule-based method, and optimal DP strategy with CBDC and true passenger load factor at 12:00 PM of Guangdong Transit Bus Line 11 on 12/15/2014: (a) power profiles of batteries, (b) power profiles of supercapacitors, (c) power demand profile, (d) SOC profiles of batteries, (e) SOC profiles of supercapacitors, and (f) total cumulated cost.

Table 7

Total costs of proposed and pure rule-based methods compare with the optimal DP cost at the peak hour (8:00 AM) and off-peak hour (12:00 PM).

| Methods | Peak Hour (8:00 AM) | Off-peak Hour (12:00 PM) |
|---|---|---|

| | Total Cost (USD) | Compare with Optimal DP | Total Cost (USD) | Compare with Optimal DP |
|---|---|---|---|---|
| Proposed method | 3.6890 | 0.0136 % more cost | 2.6869 | 0.0298 % more cost |
| Pure rule-based method | 4.1181 | 11.647 % more cost | 2.7973 | 4.1398 % more cost |

As shown as Fig. 25 (f) and Table 7, the total cost of the proposed method, pure rule-based method, and optimal DP strategy at 12:00 PM are 2.6869, 2.7973, and 2.6861 USD, respectively. In the proposed method, the V2C connectivity offers the benefit of the PLF prediction and the platform to implement a computationally expensive algorithm. As shown in the supercapacitor SOC subplot, Fig. 25 (d) and (e), the proposed method has a very similar SOC trajectory of the supercapacitors as the optimal DP strategy dose. Without charging supercapacitors beforehand, as shown in Fig. 25 (a) and (b), with the pure rule-based method, there is not enough energy stored in the supercapacitors to avoid discharging the batteries with high C-rates when power demands are high.

Because the rule of the rule-based controller of the proposed method is updated by the optimization results of the cloud DP, the rule-based controller of the proposed method is more robust to handle the different driving conditions than the rule-based controller used in the pure rule-based method. The difference in the total costs between the proposed method and the optimal DP strategy is 0.0298%, while a 25% error of the PLF is introduced. The PLF prediction error leads to errors of up to 8.936 kW in power demand prediction, and the errors between the reference power profiles and real power demands. These errors are well handled by the rule-based controller of the proposed method on the vehicle level. It also indicates that the proposed method is very robust against the prediction error of the PLF.

One thing remarkable is that, as shown as Fig. 25 (f), before the 1000 s, the cost of the pure rule-based method is lower than the costs of the proposed method and optimal DP. The reason is that with prediction and prior knowledge about the future power demand, the proposed method and rule-based method focus on minimizing the total cost of the whole circle rather than the current cost. Thus, using these two strategies, the batteries will charge and save the energy of supercapacitors for the future. It leads to higher costs at the beginning but saves more cost in the end because it allows the supercapacitors to have enough energy to supply during high power demands conditions. It further illustrates that without the predictions of future power demands, the performance of the traditional rule-based methods is limited. In addition, though the 12:00 PM is the off-peak hour, which includes the less loading effect, the proposed method leads 3.946% of cost improvement than the pure rule-based method.

4.6 Discussion the adapting of the proposed method into wider applications

Electric load prediction is a rapidly growing research area. Besides applied to city buses, the proposed passenger load prediction method can be extended to predict energy demand of other urban transportation as well, which can assist energy management systems to improve its energy efficiency. For example, the proposed passenger load prediction models can be adapted to the electric subway train. Whose weight of the car of New York City subway train, R 160 A, is about 38600 kg, and its passenger capacity is 240. Taking the average weight of a person, about 70 kg, into the calculation, the total weight change of a subway car due to the passenger load effect can be up to 30.32%. Because different PLF cause significant changes in the total weight, similar to city buses, the power demand of a subway train also highly depends on its PLF. By adopting the proposed passenger load prediction algorithm to a subway train, an accurate power demand prediction for subway trains can be realized, which is beneficial to the optimization of the safety and economic operation [35].

Meanwhile, the framework of the proposed EMS, which fully explores and utilizes the benefits of cloud techniques, can also be applied to other commercial and residential energy systems that use Li-ion batteries and supercapacitors as their energy storage sources, such as passenger cars, ships, and residential buildings. In recent papers, some innovative load prediction algorithms for these energy systems are proposed, including the Monte Carlo algorithm for electric vehicles [36], the Markov chain model for hybrid ships [37], and the bidirectional Long Short-Term Memory and Gated Recurrent Units algorithm for buildings [38]. With the framework of the proposed EMS in this paper, an optimal reference solution can be obtained by running these optimal but high time-consuming power demand prediction algorithms and DP programming on the cloud. Then, by utilizing a carefully designed rule-based controller with rule update ability, the error between the predicted and actual power demands can be handled in real-time, and the load distribution between power sources can be effectively optimized with the limited computational cost.

## 5. Conclusion

In this paper, a dynamic programming approach is utilized as the battery/ supercapacitor hybrid electric bus EMS in the battery aging effect and passenger loading effect discussion. The passenger loading effect plays a more dominant role than the battery aging effect does on the bus operation cost. About the PLF prediction, this study investigated four models, out of which GBDT model is selected to assist the vehicle EMS evaluation for its high accuracy and stability. A new EMS comprised of a cloud DP and rule-based controller with the rule update is proposed. To evaluate the proposed EMS performances, the proposed EMS is compared with the optimal DP method and the pure rule-based method. In this case, the proposed EMS and optimal DP share similar results with less than 1% difference. At peak and off-peak hours, the proposed EMS reduces 10.419% and 3.946% cost than the pure rule-based method.

In future studies, the size of the optimal components of supercapacitor and battery could be conducted. Besides, the performance of the rule-based controller used on the vehicle side can be further improved by increasing its rule with more heuristic principles in the EMS field. Furthermore, thermal safety is also critical during the operation of a hybrid electric bus [39]. More comprehensive battery models, such as the electro-thermal-aging coupling model [40], can be included in future studies to improve the practicability of the EMS.